\documentclass[twocolumn]{aastex631}
\usepackage{amsmath}

\submitjournal{ApJ}

\shorttitle{AMBER kSZ}
\shortauthors{Chen et al.}

\begin{document}

\title{Patchy Kinetic Sunyaev-Zel'dovich Effect with Controlled Reionization History and Morphology}

\correspondingauthor{Nianyi Chen}
\email{nianyic@andrew.cmu.edu}

\author{Nianyi Chen}
\affiliation{McWilliams Center for Cosmology, Department of Physics, Carnegie Mellon University, Pittsburgh, PA 15213, USA}

\author{Hy Trac}
\affiliation{McWilliams Center for Cosmology, Department of Physics, Carnegie Mellon University, Pittsburgh, PA 15213, USA}
\affiliation{NSF AI Planning Institute for Physics of the Future, Carnegie Mellon University, Pittsburgh, PA 15213, USA}
\author{Suvodip Mukherjee}
\affiliation{Perimeter Institute for Theoretical Physics, 31 Caroline Street N., Waterloo, Ontario, N2L 2Y5, Canada}
\author{Renyue Cen}
\affiliation{Department of Astrophysical Sciences, Princeton University, Princeton, NJ 08544, USA}


\begin{abstract}

Using the novel semi-numerical code for reionization AMBER, we model the patchy kinetic Sunyaev-Zel'dovich (kSZ) effect by directly specifying the reionization history with the redshift midpoint $z_\mathrm{mid}$, duration $\Delta_\mathrm{z}$, and asymmetry $A_\mathrm{z}$. 
We further control the ionizing sources and radiation through the minimum halo mass $M_\mathrm{h}$ and the radiation mean free path $\lambda_\mathrm{mfp}$. 
AMBER reproduces the free electron number density and the patchy kSZ power spectrum of radiation-hydrodynamic simulations at the target resolution ($1\,{\rm Mpc}/h$) with matched reionization parameters. 
With a suite of $(2\,{\rm Gpc}/h)^3$ simulations using AMBER, we first constrain the redshift midpoint $6.0<z_{\rm mid}<8.9$ using the Planck2018 Thomson optical depth result (95\% CL). 
Then, assuming $z_{\rm mid}=8$, we find that the amplitude of $D^{\rm pkSZ}_{\ell=3000}$ scales linearly with the duration of reionization $\Delta_z$, and is consistent with the $1\sigma$ upper limit from the South Pole Telescope (SPT) results up to $\Delta_z<5.1$ ($\Delta_z$ encloses $5\%$ to $95\%$ ionization). 
Moreover, a shorter $\lambda_{\rm mfp}$ can lead to a $\sim 10\%$ lower $D^{\rm pkSZ}_{\ell=3000}$ and a flatter slope in the $D^{\rm pkSZ}_{\ell=3000}-\Delta_z$ scaling relation, thereby affecting the constraints on $\Delta_z$ at $\ell=3000$. 
Allowing $z_{\rm mid}$ and $\lambda_{\rm mfp}$ to vary simultaneously, we get spectra consistent with the SPT result ($95\%$ CL) up to $\Delta_z=12.8$ (but $A_z>8$ is needed to ensure an end of reionization before $z=5.5$). We show that constraints on the asymmetry require $\sim 0.1\,\mu k^2$ measurement accuracy at multipoles other than $\ell=3000$. 
Finally, we find that the amplitude and shape of the kSZ spectrum are only weakly sensitive to $M_h$ under a fixed reionization history and radiation mean-free path.
\end{abstract}


\section{Introduction}
\label{sec:introduction}

The epoch of reionization (EoR) is the period in cosmic history when ionizing radiation emitted by the first galaxies and quasars ionized the baryons in the Universe, leading to a transition of the gas content from a neutral state to an ionized state.
Because EoR happens at a relatively high redshift ($z =5\sim 15$), the limited observational evidence has hindered our full understanding of the whole physical process involved.
Nevertheless, recent and future experiments using various probes are making the picture of the EoR more and more complete. 
For example, \citet{Planck2018arXiv180706209P} recently inferred $\tau =0.054\pm 0.007$ from measurements of the Cosmic Wave Background (CMB) temperature and polarization angular power spectra, implying a late reionization midpoint at redshift $z \approx 7.7 \pm 0.6$ \citep[e.g.][]{Glazer2018RNAAS...2c.135G}. \citet{Becker2015MNRAS.447.3402B} find evidence of a dark Ly$\alpha$ trough extending down to $z \approx 5.5$ in the spectrum of a high-redshift quasar, suggesting that reionization could have ended at $z < 6$ \citep[e.g.][]{Keating2019arXiv190512640K}, later than previously assumed.
In addition, we also expect to gain tomographical information of the EoR through the 21cm observations such as Hydrogen Epoch of Reionization Array \citep[][]{HERA2017} and Square Kilometer Array \citep[][]{SKA2015}, and a better understanding of the first ionizing sources through the space-based telescopes such as the James Webb Space Telescope \citep[][]{JWST2006} and Roman Space Telescope \citep[][]{WFIRST2015}.

Of particular interest to this paper is the use of CMB secondary at high multipoles to constrain the EoR. With the improvements in recent ground-based CMB experiments such as  Atacama Cosmology Telescope (ACT\footnote{\url{https://act.princeton.edu}}) and the South Pole Telescope (SPT\footnote{\url{http://pole.uchicago.edu}}), we are already able to use anisotropies in the CMB temperature map to constrain reionization through Sunyaev-Zel'dovich (SZ) effect \citep{Zeldovich1969,Sunyaev1980ARA&A..18..537S}. 
The SZ effect results from inverse-Compton scattering of CMB photons off high-energy electrons in the IGM, and it has the largest contribution among CMB secondary anisotropies on arc-minute scales. 
There are two types of SZ effect: thermal SZ effect (tSZ) comes from the electron pressure within the intra-cluster medium (ICM) and has a spectrum shifted from the CMB black body spectrum, while kinetic SZ (kSZ) effect is due to the bulk motion of electrons in the IGM with respect to the CMB rest frame and has the same spectrum as the CMB \citep[e.g.][]{Carlstrom2002}. 
kSZ signal can be further divided into two components and they have comparable amplitude \citep[e.g.][]{Trac2011,Shaw2012}: patchy kSZ originates from the inhomogeneous free-electron fraction in the universe during cosmic reionization, and homogeneous kSZ results from the peculiar velocities of the galaxies after the universe is fully ionized \citep[e.g.][]{Ostriker1986ApJ...306L..51O}.

Because patchy kSZ originates from inhomogeneous reionization, its amplitude and power spectrum are sensitive to the timing, duration, and detailed history of reionization. 
Thus, by probing patchy kSZ fluctuation we can put constraints on reionization history provided that we have a thorough understanding of their relation. 
In recent years, developments in numerical simulations enables us to understand connection between the patchy kSZ angular power spectrum and reionization \citep[e.g.][]{Zhang2004,McQuinn2005,Iliev2007ApJ...660..933I,Tashiro2011,Mesinger2011MNRAS.411..955M,Battaglia2013ApJ...776...83B,Park2013ApJ...769...93P,Alvarez2016ApJ...824..118A,Gorce2020A&A...640A..90G,Choudhury2021,Paul2021}.
In particular, multiple works have demonstrated that semi-numerical simulations are powerful tools to study kSZ with various reionization scenarios in a relatively quick fashion.
For example, \citet{Mesinger2012MNRAS.422.1403M} and \citet{Choudhury2021} have used the semi-numerical simulations to study the dependence of reionization history and patchy kSZ power spectrum on the ionizing efficiency of high-redshift galaxies, the minimum virial temperature of haloes, and the ionizing photon mean free path.
\citet{Battaglia2013ApJ...776...83B} combined N-body simulations with post-processed reionization-redshift field to study the effect of reionization history on the patchy kSZ power spectrum.
\citet{Alvarez2016ApJ...824..118A} uses very large-scale simulations to study the different components to the kSZ signal, as well as the four-point statistics of patchy kSZ.

However, the majority of semi-numerical codes of reionization are based on the excursion set formalism method for reionization \citep[e.g.][]{Bond1991ApJ...379..440B,Furlanetto2004ApJ...613....1F}, and it has been shown that \citep{Zahn2011MNRAS.414..727Z} these semi-numerical methods are in good agreement with radiative transfer simulations when compared at the same ionization fraction, but not at the same redshift without renormalization. 
Moreover, most of the semi-analytical models parametrize reionization on the power-spectrum level \citep[e.g.][]{Shaw2012,Battaglia2013ApJ...776...83B,Gorce2020A&A...640A..90G}, without directly controlling the reionization history \citep[but see ][who controls the duration of reionization by varying the ionizing efficiency across redshift]{Paul2021}.
This motivates us to study the patchy kSZ signal using the novel semi-numerical simulation Abundance Matching Box for the Epoch of Reionization \citep[AMBER;][]{Trac2021}, which takes reionization history as a direct input.

In this paper, we use the semi-numerical simulation AMBER to generate reionization CMB observables such as the Thomson optical depth and patchy kSZ for different sets of reionization history parameters and cosmological parameters. 
By doing so, we can disentangle the effect of individual parameters on the observed spectra. 
We present the dependence of kSZ power spectrum and Thomson optical depth on reionization parameters as well as cosmological parameters.

The paper is organized as follows:
In Section \ref{sec:tau_ksz}, we introduce the theory and computation of the Thompson optical depth, patchy kSZ effect, and the patchy kSZ angular power spectra.
In Section \ref{sec:sim}, we summarize the semi-analytical models used in AMBER, as well as the RadHydro simulations we use to calibrate the AMBER models.
Section \ref{sec:comparison} shows comparisons between the AMBER outputs and the RadHydro simulations with matched reionization parameters and resolutions.
In Section \ref{sec:params}, we systematically study the effect of reionization parameters and cosmological parameters on the patchy kSZ signal, including the maps and angular power spectra, and compare the results with observational constraints.
Unless otherwise stated, we assume a flat $\Lambda$CDM cosmology with $[\Omega_m,\Omega_b,\sigma_8,n_s,h]=[0.3,0.045,0.8,0.96,0.7]$, and our fiducial values for reionization parameters are $[z_{\rm mid},\Delta_z,A_z,M_h(M_\odot),\lambda_{\rm mfp}({\rm Mpc}/h)]=[8.0,4.0,3.0,10^8,3.0]$ (see later sections for a detailed description of these parameters).

\section{Thomson Optical Depth and kSZ Effect} \label{sec:tau_ksz}

\subsection{Thomson Optical Depth} \label{sec:tau}

In the ongoing CMB experiment, the most well-constrained EoR observable has been the Thomson-scattering optical depth $\tau_e$.
Constraints on $\tau_e$ informs us about the integrated electron number density $n_e$ along the light-of-sight, as $\tau_e$ is related to $n_e$ by:
\begin{equation}
\label{eq:tau}
    \tau_e(z,\hat{\bold{n}}) = \sigma_T \int_0^{z}  dz' \frac{cdt}{dz'} n_e(z',\hat{\mathbf{n}}),
\end{equation}
where $\sigma_T$ is the Thomson scattering cross section, $\hat{\bold{n}}$ is the direction of observation, and $n_e(z',\hat{\bold{n}})$ is the free electron number density at a specific redshift in the observed direction.
The angular variation in $\tau_e$ is weak, so usually we drop the angular dependence in the computation, and use the global ionization fraction to compute $\tau_e(z)$ instead:
\begin{equation}
    \tau_{\rm{e}}(z) 
    = \sigma_T \int_0^{z}  dz' \frac{cdt}{dz'} \langle n_{\rm e}(z') \rangle_{\rm V} ,
\end{equation}
where $ \langle n_{\rm e}(z') \rangle_{\rm V}$ is the volume-averaged free electron number density.

Because the mean baryon number density increases with redshift, an earlier reionization leads to higher values of $\tau_e$, and it has been shown that $\tau_e$ is not sensitive to the detailed reionization histories beyond the redshift \citep[e.g.][]{Battaglia2013ApJ...776...83B}.
Moreover, current constraints on $\tau_e$ are primarily driven by the measurement of the low-$\ell$ EE polarization power spectrum (which is proportional to $\tau_e^2$), and is independent of the small-scale anisotropies (in particular the patchy kSZ effect). 
Therefore, constraints on $\tau_e$ help break the degeneracy of reionization history parameters in the small-scale patchy kSZ measurements.
Reversely, one can also use the patchy kSZ signal to break the degeneracy between $\tau_e$ and the primordial amplitude of scalar fluctuations $A_s$ \citep[e.g.][]{Alvarez2021}.

\subsection{Patchy kSZ Effect} \label{sec:ksz}

Next, we introduce the patchy kSZ effect and the computation of the kSZ power spectrum. 
The patchy kSZ effect is the temperature fluctuation in the CMB due to the scattering of CMB photons off of free electrons in bulk motion during cosmic reionization. 
Small-scale temperature anisotropies are then generated by the coupling of large-scale velocity perturbations and the patchiness of the ionized field on small scales.
The fractional temperature fluctuation induced by patchy kSZ effect is calculated by integrating the electron momentum along the line of sight:
\begin{equation}
\label{eq:ksz}
    \frac{\Delta T_{ksz}}{T} =  \sigma_T \int dz \frac{cdt}{dz} e^{-\tau_e(z,\hat{\bold{n}})} n_e(z,\hat{\bold{n}})  \hat{\bold{n}}\cdot \bold{v}
\end{equation}
where $\bold{v}$ is the peculiar velocity of free electrons, $\tau_e$ is the Thomson optical depth calculated in Eq.\ref{eq:tau}, and the integration limits are the beginning and end of reionization.

Currently, most constraints on reionization from patchy kSZ come from the angular power spectrum \citep[but see e.g.][for the use of high-order statistics]{Smith}. This is also the quantity of interest in this paper. 
To compute the kSZ angular power spectrum, we use the Limber approximation following \cite{Park2013ApJ...769...93P}. 
First, we define the specific free electron momentum as:
\begin{equation}
    \mathbf{q} = x_i \mathbf{v} (1+\delta),
\end{equation}
where $x_i$ is the mass-weighted ionization fraction, $\mathbf{v}$ is the peculiar gas velocity, and $\delta$ is the gas overdensity. Then, Equation \ref{eq:ksz} can be re-written as:
\begin{equation}
    \label{eq:ksz_limber}
     \frac{\Delta T_{ksz}}{T}(\hat{\bold{n}}) =  \sigma_T n_{e,0}\int_{z_{beg}}^{z_{end}} \frac{ds}{a^2} e^{-\tau(z)}  \hat{\bold{n}}\cdot \bold{q},
\end{equation}
where $n_{e,0}$ is the total number of electrons at the present epoch, $a$ is the scale factor, and $s$ is the comoving distance. 
The kSZ angular power spectrum is given by:
\begin{equation}
    C_{\ell} = \left(\frac{\sigma_T n_{e,0}}{c}\right)^2 \int \frac{ds}{s^2 a^4} e^{-2\tau(z)} \frac{P_{q_\perp} (k=\ell/s,s)}{2}.
\end{equation}
To compute $P_{q_\perp} (k)$, let $\Tilde{\mathbf{q}}$ be the 3D Fourier transform of the momentum field. Then, the projection of $\Tilde{\mathbf{q}}$ on the plane perpendicular to the mode vector is given by $\Tilde{\mathbf{q}}_\perp(\mathbf{k}) = \Tilde{\mathbf{q}}(\mathbf{k}) - \hat{k}(\Tilde{\mathbf{q}}(\mathbf{k}) \cdot \hat{k})$. Finally, $P_{q_\perp} (k)$ is the power spectrum of $\Tilde{\mathbf{q}}_\perp(\mathbf{k})$ given by:
\begin{equation}
    (2\pi)^3 P_{q_\perp} (k) \delta(\mathbf{k} - \mathbf{k'}) = \langle \Tilde{\mathbf{q}}_\perp(\mathbf{k}) \cdot \Tilde{\mathbf{q}}^*_\perp(\mathbf{k}) \rangle.
\end{equation}
Note that we can also directly compute the kSZ power spectrum from full-sky patchy kSZ maps. 
However, since the scale of interest is small ($\ell>1000$) and is well-approximated by the Limber approach, we choose to follow the Limber approximation for a less noisy spectrum and faster computation.

\section{Simulations} \label{sec:sim}
The simulations in this work are run with the new semi-numerical code AMBER \citep{Trac2021}.
In this section, we will introduce the main models in the AMBER code relevant for calculating the kSZ signal. 
Moreover, we also briefly introduce the RadHydro simulation suite from the Simulations and Constructions of the Reionization of Cosmic Hydrogen (SCORCH) project \citep[][]{Trac2015ApJ...813...54T,Doussot2019ApJ...870...18D,Chen2020ApJ...905..132C}.
We will later use these RadHydro simulations to calibrate AMBER models and compare results from both simulations.

\subsection{AMBER} \label{subsec:amber}

\subsubsection{Parametrization of the Reionization History}

\begin{figure}[t]
\includegraphics[width=\hsize]{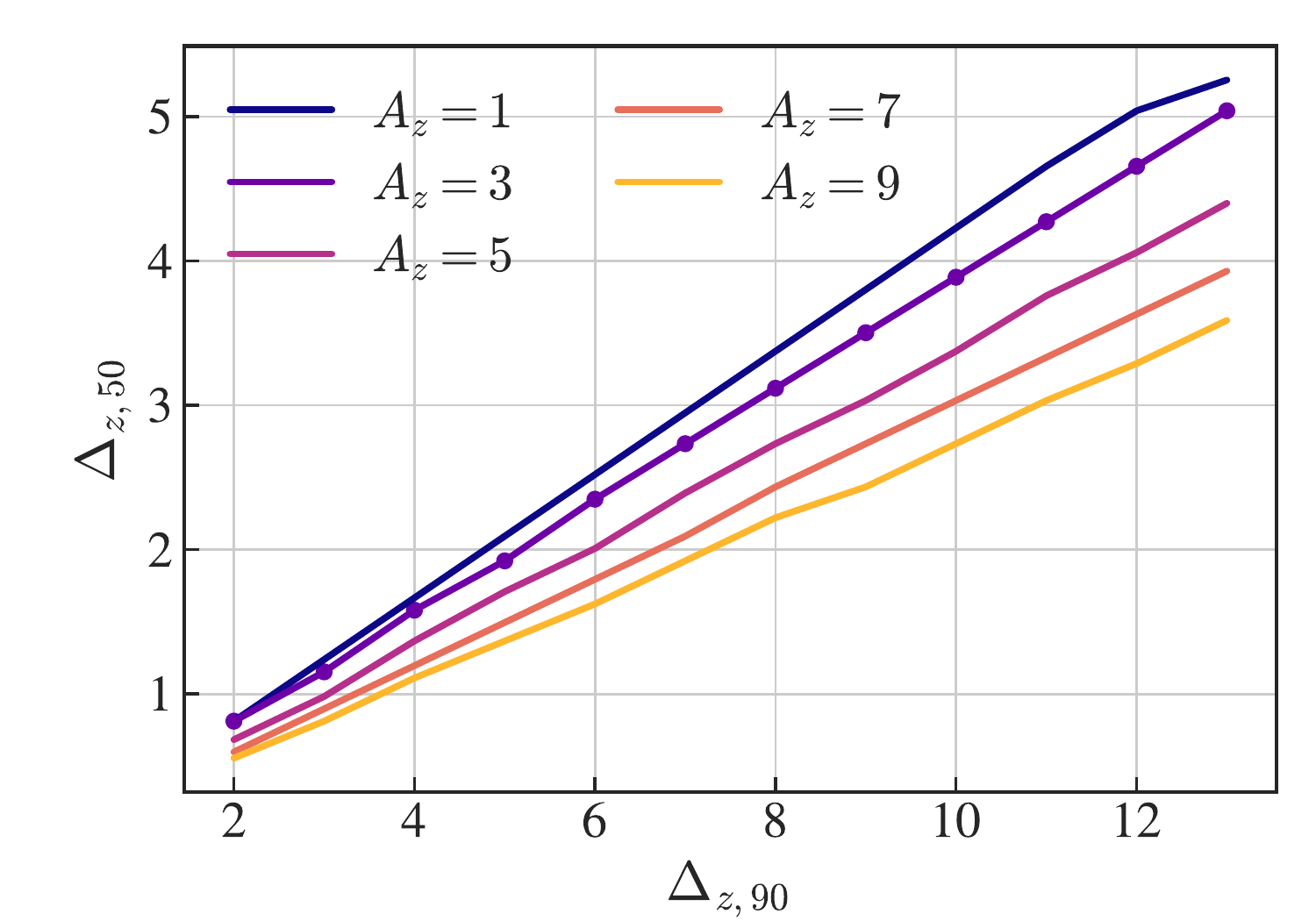}
\caption{Scaling relation between $\Delta_{z,90}$ and $\Delta_{z,50}$ for different levels of asymmetry.}
\label{fig:d50_d90}
\end{figure}

The reionization history $x_\mathrm{i}(z)$ is the fraction of hydrogen that are ionized at a certain redshift.  
It is of primary interest to our understanding of the EoR, because it reveals the possible astrophysical process during the EoR, and also directly affects many key EoR observables. 
For example, the integrated Thomson optical depth and the evolution of the global 21cm brightness temperature depend linearly on the ionized electron fraction $\bar{x}_\mathrm{i}(z)$ and neutral hydrogen fraction $\bar{x}_\mathrm{HI}(z)$, respectively. 
One of the major novelties of AMBER is that we directly control the reionization history in our reionization modeling.

Following the argument in \cite{Trac2008ApJ...689L..81T}, AMBER parametrizes the reionization history with the midpoint, duration, and asymmetry parameters. In AMBER, we use the mass-weighted global ionization fraction $\bar{x}_\mathrm{i,M}$, and drop the subscript hereafter. 

We define the midpoint redshift $z_\mathrm{mid}$ as the redshift at which $50\%$ of the universe is ionized by mass.
To characterize the duration of reionization $\Delta_z$, let $z_{\rm ear} > z_\mathrm{mid}$ and $z_{\rm lat}<z_\mathrm{mid}$ correspond to the early and late stage of reionization, respectively. 
The duration is then defined as:
\begin{equation}
\label{eqn:duration}
\Delta_\mathrm{z} \equiv z_\mathrm{ear} - z_\mathrm{lat}.
\end{equation}
There are various definitions of $\Delta_z$ in previous works depending on how one defines $z_{\rm ear} $ and $ z_{\rm lat}$.
Throughout this work, we take $(z_{\rm ear},z_{\rm lat})$ as the redshifts at which the universe is $5\%$ and $95\%$ ionized, respectively.
Under this definition, $\Delta_\text{z}$ more effectively quantifies the whole EoR. 
We refer to this definition as $\Delta_{z,90}$.
We will drop the subscript "90" when we are not explicitly comparing with other definitions of $\Delta_z$.
Another popular definition takes $(z_{\rm ear},z_{\rm lat})$ as the redshifts at which the universe is $25\%$ and $75\%$ ionized \citep[e.g.][]{Battaglia2013ApJ...776...81B,Gorce2020A&A...640A..90G}. We will refer to the duration under this definition as $\Delta_{z,50}$, as it encloses $50\%$ of the ionization process.

Finally, to characterize the likely asymmetric reionization scenarios, we define the asymmetry parameter as:
\begin{equation}
\label{eqn:asymmetry}
A_\mathrm{z} \equiv \frac{z_\mathrm{ear} - z_\mathrm{mid}}{z_\mathrm{mid} - z_\mathrm{lat}} .
\end{equation}
Symmetric reionization histories would correspond to $A_\mathrm{z} = 1$, but reionization simulations typically find that the early stage of reionization takes longer than the later stage such that $A_\mathrm{z} > 1$. 
Note that with different levels of asymmetry, there is not a one-to-one correspondence between $\Delta_{z,90}$ and $\Delta_{z,50}$.
In Figure \ref{fig:d50_d90}, we show the relation between $\Delta_{z,90}$ and $\Delta_{z,50}$ for asymmetries ranging from 1 to 9. 
When the asymmetry level is lower, $\Delta_{z,90}$ corresponds to a higher value of $\Delta_{z,50}$. 
Hence, the scaling coefficient between observables and $\Delta_{z,50}$ is also affected by $A_z$.
This is important for interpreting the comparisons between our results and previous works in later sections.


In AMBER, we interpolate the three ionization points at ($z_\mathrm{ear},z_\mathrm{mid},z_\mathrm{lat}$) with a modified Weibull function \citep{Weibull1951JAM....18..293W},
\begin{equation}
    \label{eqn:weibull}
    \bar{x}_\mathrm{i}(z) = \exp\left[-\max\left(\frac{z-a_\mathrm{w}}{b_\mathrm{w}},0\right)^{c_\mathrm{w}}\right] ,
\end{equation}
where the coefficients $a_\mathrm{w}$, $b_\mathrm{w}$, $c_\mathrm{w}$ are all positive values. 
The coefficients can be easily determined by first solving a nonlinear equation for $c_\mathrm{w}$ and then substituting its value into algebraic equations for the other two coefficients. 
We find that solutions exist for the asymmetry range $A_\mathrm{z} \lesssim 15$, which is more than sufficient for parameter space studies.

\begin{deluxetable*}{lCCCCCCCCCC}
\label{tab:scorch}
\tablewidth{\textwidth}
\tablecaption{Simulation parameters and measured reionization history parameters for the three RadHydro simulations in SCORCH II.}
\tablehead{
\colhead{Model} & \colhead{$L\ [h^{-1}\text{Mpc}]$} & \colhead{$N_\text{dm}$} & \colhead{$N_\text{gas}$} & \colhead{$N_\text{RT}$} &
\colhead{$f_8$} & \colhead{$a_8$} & \colhead{$\tau$} & \colhead{$z_\text{mid}$} &  \colhead{$\Delta_\text{z}$} &  \colhead{$A_\text{z}$}
}
\startdata
Sim 0 & 50 & $2048^3$ & $2048^3$ & $512^3$ & 0.15 & 0 & 0.060 & 7.95 & 4.68 & 2.90 \\
Sim 1 & & & & & 0.13 & 1 & 0.060 & 7.91 & 5.45 & 2.69 \\
Sim 2 & & & & & 0.11 & 2 & 0.060 & 7.83 & 6.54 & 2.33
\enddata
\end{deluxetable*}
\subsubsection{Reionization-Redshift Field}

The key assumption in AMBER is that the order in which a cell gets ionized is determined by the relative radiation intensity in that cell.
In this way, given a global reionization history $ \bar{x}_\mathrm{i}(z)$, we can obtain a reionization redshift field by abundance matching against the unnormalized radiation intensity.

To begin with, the dark matter density and velocity fields are generated with second-order perturbation theory (2LPT) at the desired redshift.
On moderately nonlinear scales, the dark matter and gas distributions are highly correlated and assumed to exactly trace each other. 
Thus we use the dark matter overdensity to approximate the gas overdensity in AMBER.

We then construct halo mass density fields with the Lagrangian version of the excursion set formalism (ESF-L) \citep[see][for more detailed descriptions]{Trac2021}. 
We use the minimum halo mass parameter $M_h$ to control the lowest halo mass for hosting ionizing sources.
This step gives us the halo density field $\rho_\mathrm{halo}(\boldsymbol{x})$, which is a proxy of the ionizing sources in the simulation.

Then, assuming that radiation intensity of ionizing sources is proportional to the halo density, and that the photon flux attenuation follows $e^{-r/\lambda_\mathrm{mfp}}$, we obtain the (unnormalized) radiation intensity field $r(\boldsymbol{x},z)$ by convolving $\rho_\mathrm{halo}(\boldsymbol{x})$ with a kernel function $\frac{1}{4\pi r^2}\exp\left(-\frac{r}{\lambda_\mathrm{mfp}}\right)$.
$\lambda_{\text{mfp}}$ would affect how much radiation is received by each cell in our simulation, and as a result how early each cell is ionized. 
Here we use an effective mean free path $\lambda_{\rm mfp}$ to account for the attenuation of the radiation field. 
In principle, the mean free path of photons is a local variable that could depend on the halo mass and redshift.
However, given the semi-analytical nature of our model and the resolution at $1 \;{\rm Mpc}/h$, we set the photon mean free path as a global variable $\lambda_{\text{mfp}}$. 
We plan to incorporate the temporal and spatial variations of $\lambda_{\text{mfp}}$ in future developments.

Finally, the reionization-redshift field $z_\mathrm{re}(\boldsymbol{x})$ is assumed to be correlated with the radiation field $r(\boldsymbol{x},z)$.
A region with higher radiation intensity is considered to be photoionized earlier and has a higher reionization-redshift. 
The abundance matching technique assigns redshift values such that the reionization history follows a given mass-weighted ionization fraction $\bar{x}_\mathrm{i}(z)$, specified with the redshift midpoint, duration, and asymmetry parameters and interpolated with a Weibull function (Equation \ref{eqn:weibull}).

We perform the abundance matching based on the radiation field at a single redshift $z_\mathrm{mid}$ for computational efficiency, but it can also be done tomographically using multiple redshift intervals. At a given redshift bin $z_n$, we have a corresponding mass-weighted ionization fraction $\bar{x}_\mathrm{i}(z_n)$ from the specified reionization history. 
We then rank order the cells at this redshift in descending order by $r(\boldsymbol{x},z_\mathrm{mid})$.
Then we ionize the first $k_n$ cells by this rank such that we reach an ionized mass fraction of $\bar{x}_\mathrm{i}(z_n)$.
Here the ionized mass fraction is calculated from the linearly extrapolated overdensity with respect to the overdensity at the midpoint redshift.
We note that the volume-weighted ionization fraction in this case is $k_n/N_{\rm cell}$ and it is typically lower than the mass-weighted $\bar{x}_\mathrm{i}(z_n)$.

\subsection{RadHydro} \label{sec:scorch}

\begin{figure*}[t]
\centering
\includegraphics[width=0.88\textwidth]{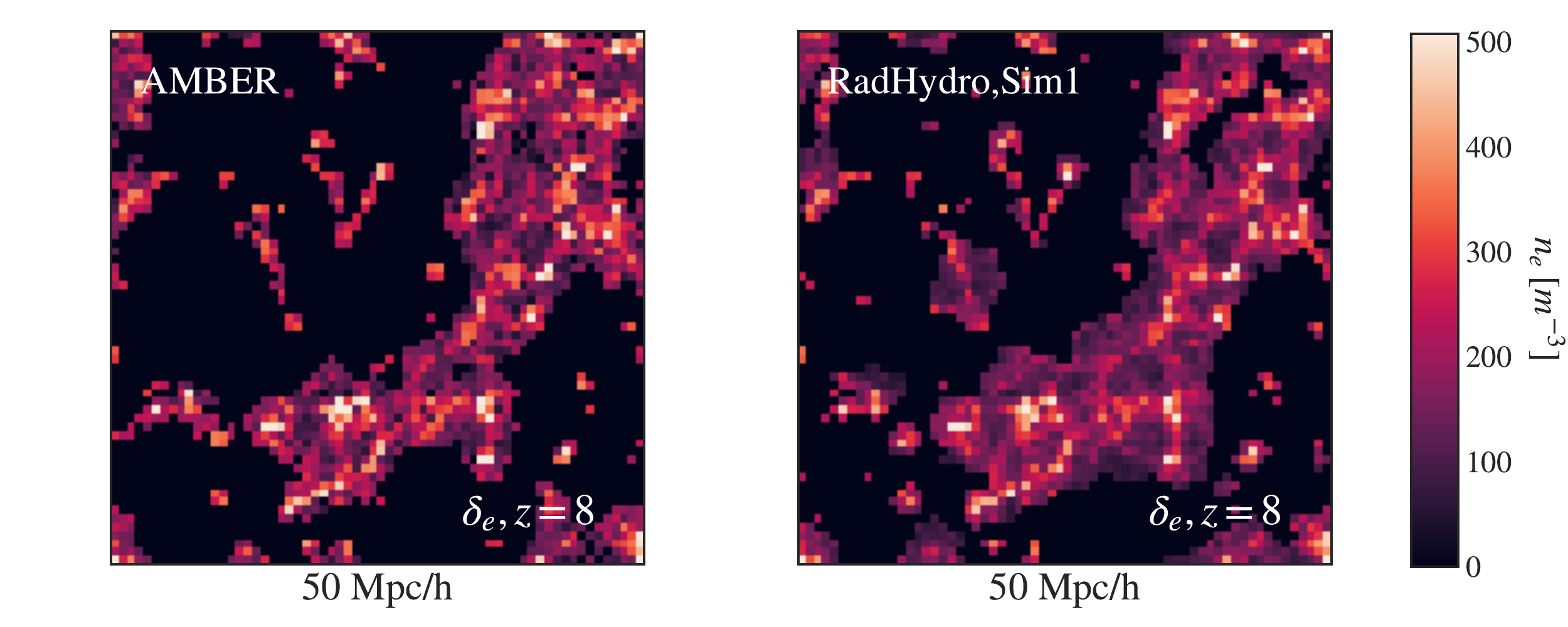}
\caption{Visualization of a $50\;{\rm Mpc}/h\times 50\;{\rm Mpc}/h\times 1\;{\rm Mpc}/h$ (shown in $64^2$ pixels) slice of the free electron number density at $z=8$, from the RadHydro Sim 1 and AMBER with the matched reionization parameters. AMBER produce slightly more concentrated ionized regions compared to RadHydro.} 
\label{fig:ne_visual}
\end{figure*} 

The Simulations and Constructions of the Reionization of Cosmic Hydrogen (SCORCH) project \citep[][]{Trac2015ApJ...813...54T,Doussot2019ApJ...870...18D,Chen2020ApJ...905..132C} is a set of N-body and radiation-hydrodynamic simulations that is designed to provide theoretical predictions and mock observations of reionization for more accurate comparisons with present and future observations. 
It is the motivation of AMBER, so we will briefly summarize the SCORCH simulations here. For details of the SCORCH project, please refer to  \citet{Trac2015ApJ...813...54T} and \citet{Doussot2019ApJ...870...18D}.

SCORCH II \citep{Doussot2019ApJ...870...18D} is a set of three radiation-hydrodynamic (RadHydro) simulations with the same cosmic initial conditions, same galaxy luminosity functions, but with different radiation escape fraction $f_{\rm esc}(z)$ models. 
The simulations are designed to have the same Thomson optical depth $\tau \approx 0.06$, consistent with recent CMB observations \citep{Planck2018arXiv180706209P}, and similar midpoints of reionization $7.5 \lesssim z \lesssim 8$, but with different evolution of the ionization fraction $\bar{x}_\text{i}(z)$. 
They model high-redshift galaxies using an updated subgrid approach that allows systematically control of the galaxy distributions in the simulations while matching the observed luminosity functions from HST \citep[e.g.][]{Bouwens2015ApJ...803...34B,Finkelstein2015ApJ...810...71F}. 

Table \ref{tab:scorch} summarizes the parameters for the three RadHydro simulations, as well as the measured midpoint, duration, and asymmetry parameters from the simulations.

\section{Calibration and Comparison with RadHydro}
\label{sec:comparison}

\begin{figure}[t]
\includegraphics[width=0.5\textwidth]{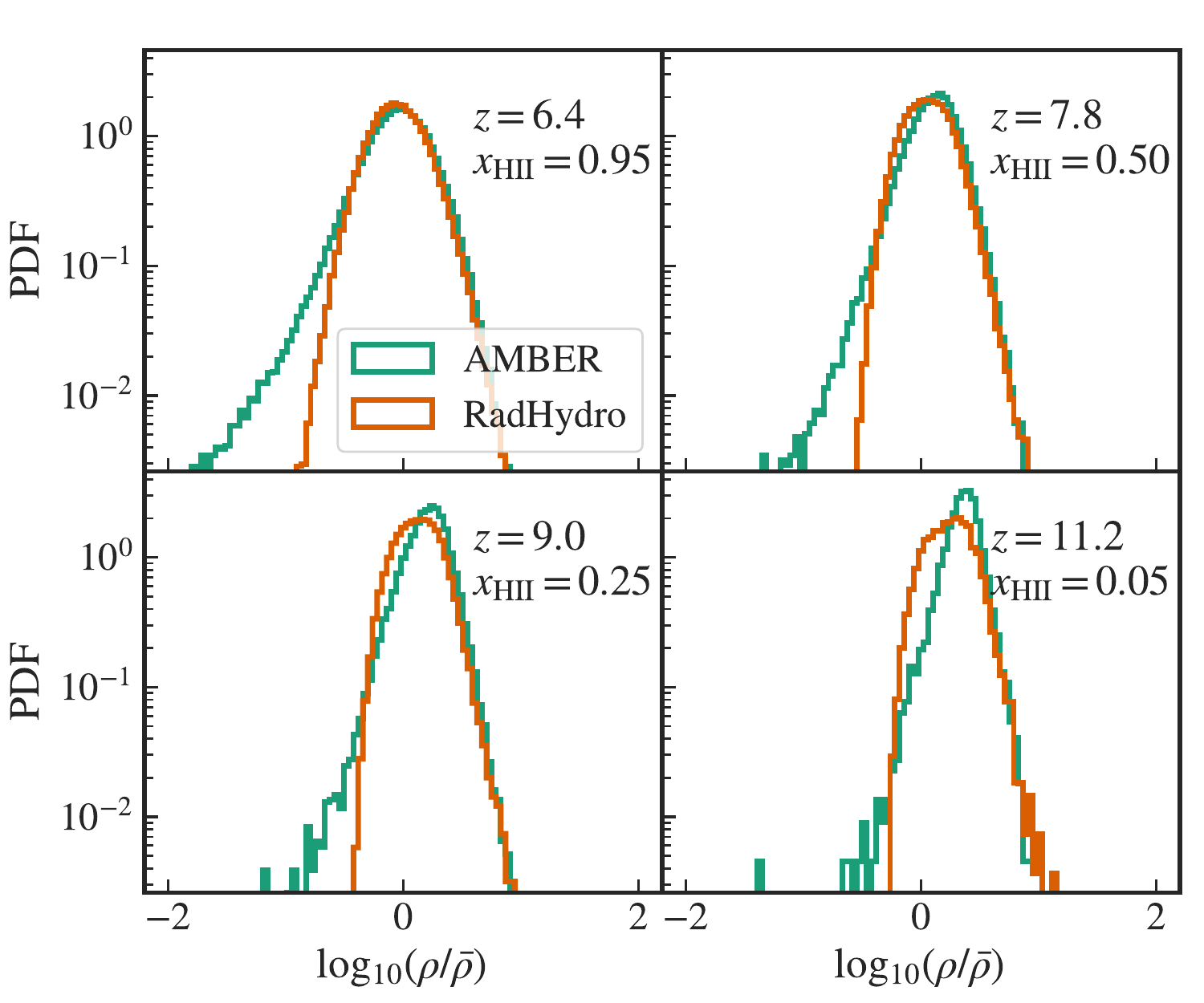}
\caption{The distribution of gas densities in ionized regions. The ionized regions in AMBER have a peak at higher densities compared to RadHydro, especially during the early stage of reionization ($\bar{x}_{i}<0.5$).  The disagreement in under-dense regions is not concerning because it is due to the much smaller number of LPT particles and the different assignment and deconvolution process for AMBER. We would find the same effects for RadHydro if we use a lower resolution simulation and not done the simple binning.} 
\label{fig:ne_pdf}
\end{figure}

\begin{figure}[t]
\includegraphics[width=0.5\textwidth]{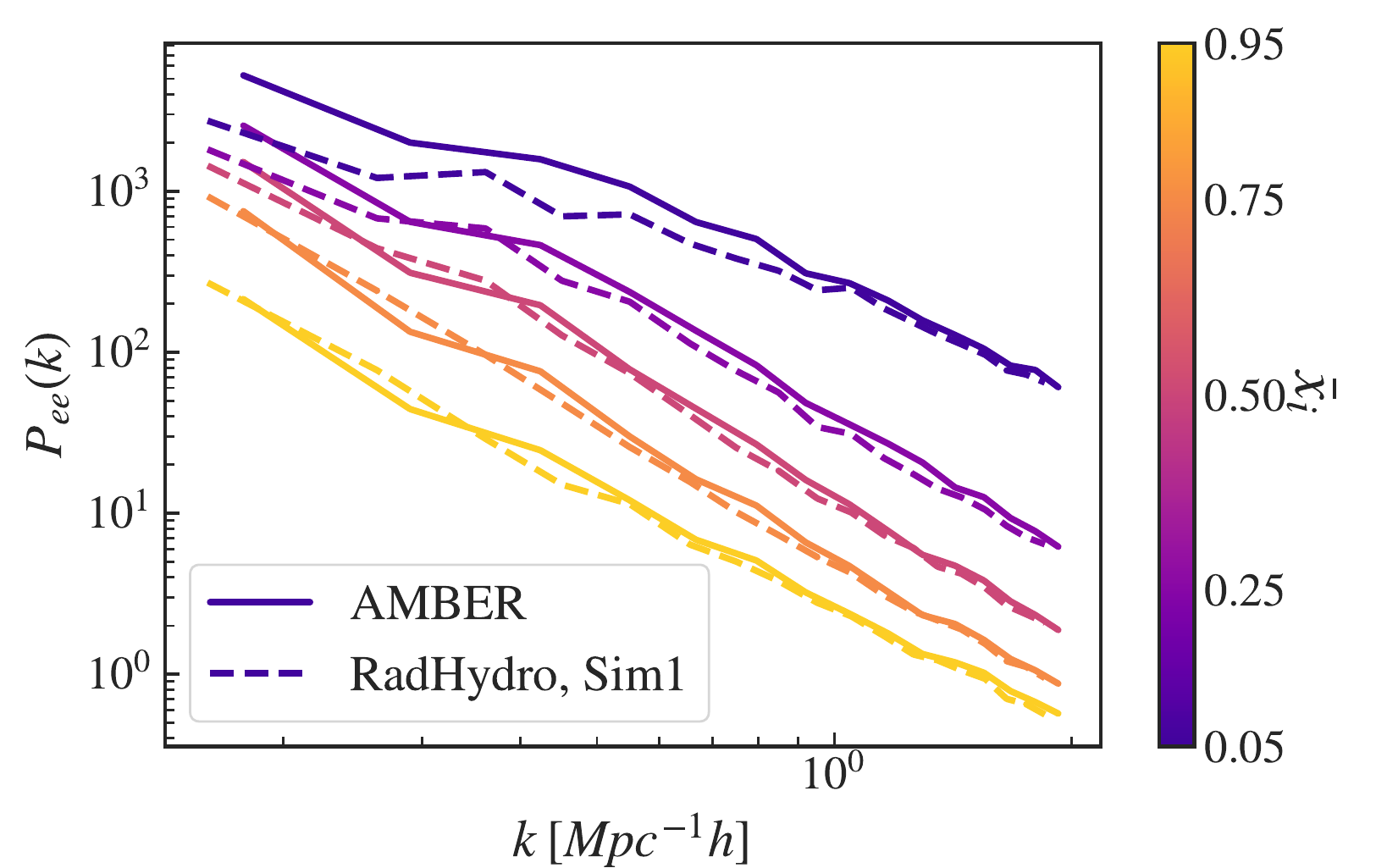}
\caption{Comparison between the AMBER and RadHydro free electron number density power spectraum $P_{ee}(k,z)$ at different redshifts. In both simulations, the overall $P_{ee}(k,z)$ gets lower as reionization evolves, because as more mass in low-density regions get ionized, the ionized electron field becomes a less biased tracer of the matter density field. At very high redshift, AMBER has larger power on large scales, but from $z=9$ onwards, the two simulation matches well with each other.} 
\label{fig:ne_power}
\end{figure} 

\begin{figure*}[t]
\centering
\includegraphics[width=0.9\textwidth]{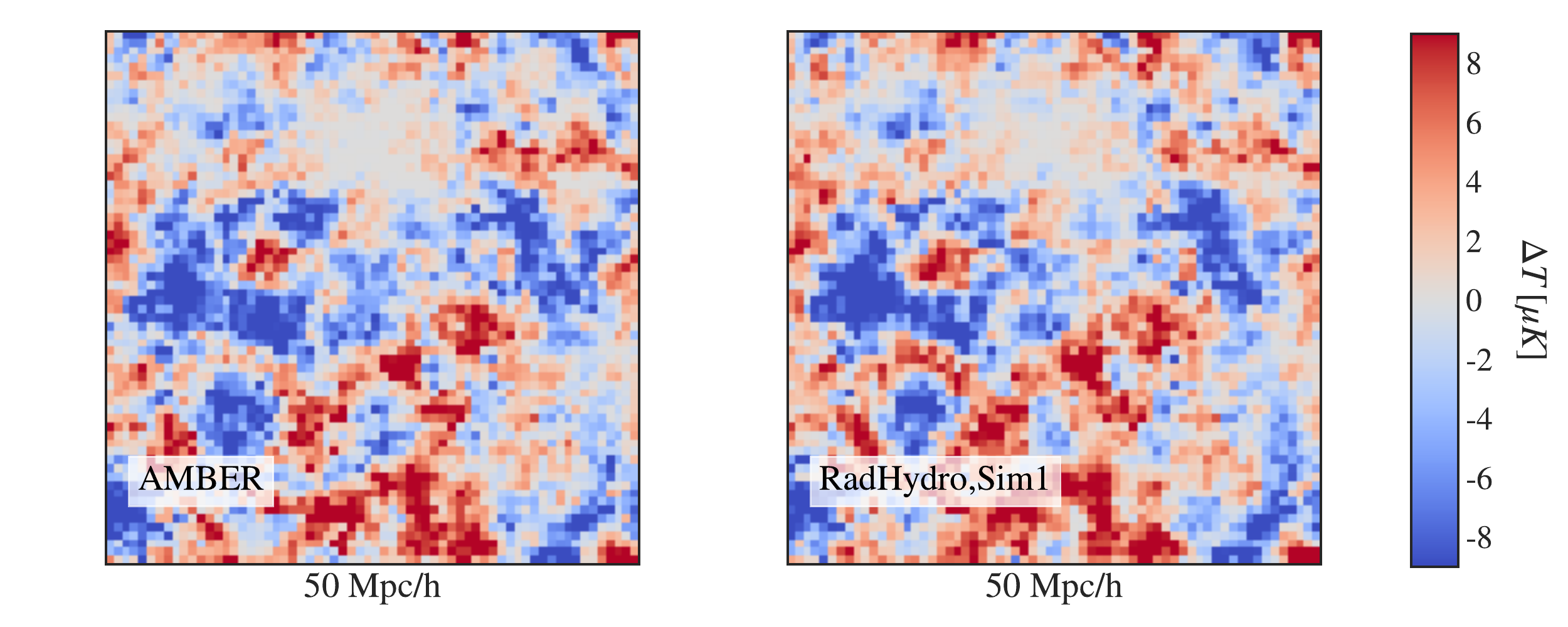}
\caption{The 2D projected patchy kSZ temperature maps under flat sky approximation for RadHydro Sim 1 (\textit{right}) and AMBER (\textit{left}) in the same simulation as described in Figure \ref{fig:ne_visual}. The projection is done at the redshift $z=8.0$. Here we sum along $z$ axis the electron momentum to get the fractional temperature difference in the CMB. The projected kSZ map of AMBER resembles that of RadHydro Sim 1 when the reionization parameters are matched.}
\label{fig:compare_ksz_map}
\end{figure*} 

\begin{figure}[t]
\centering
\includegraphics[width=0.49\textwidth]{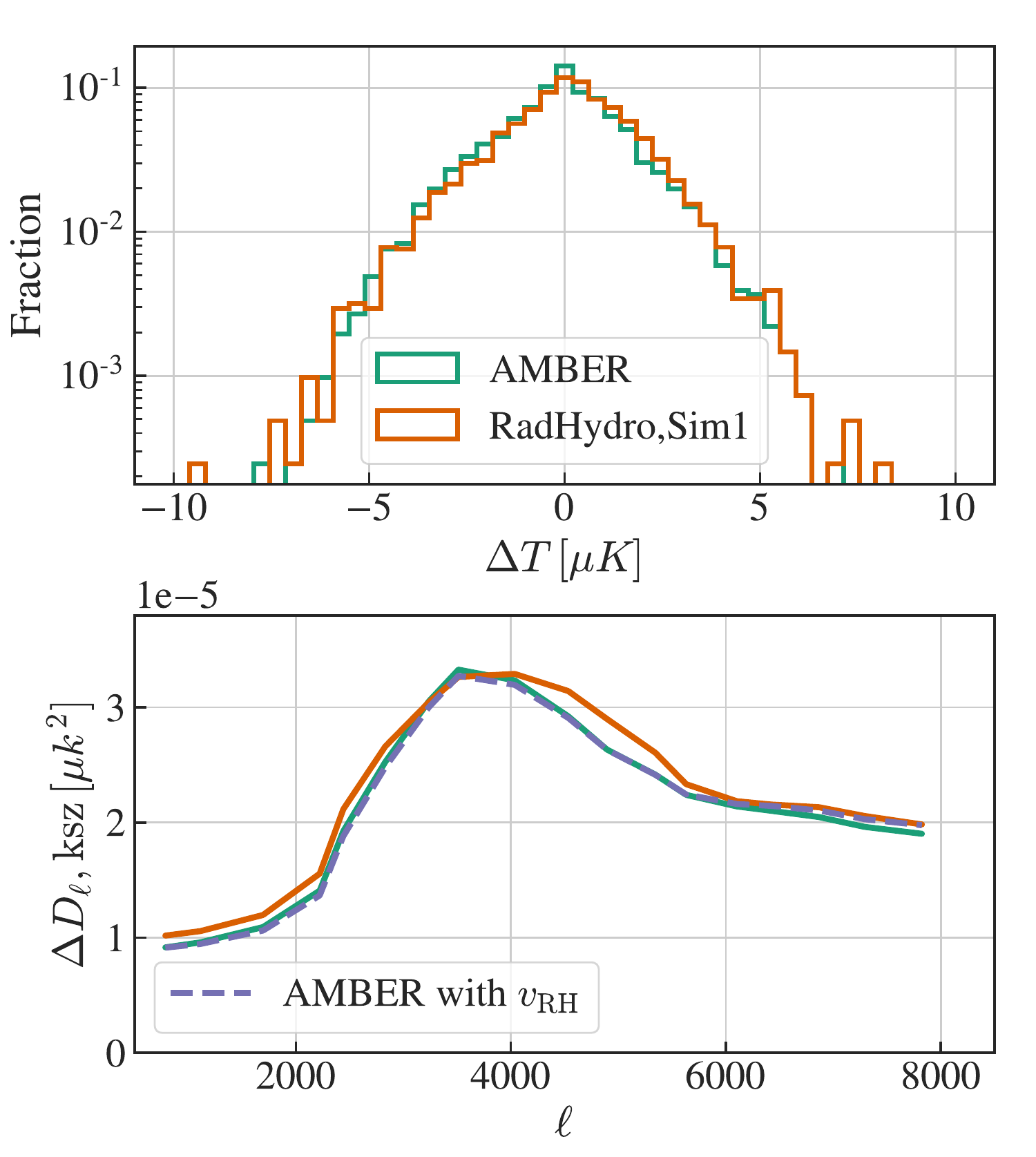}
\caption{\textbf{\textit{Top:}} The distribution of $\Delta T$ of the projected kSZ maps generated using AMBER (\textit{green}) and RadHydro (\textit{orange}) shown in Figure \ref{fig:compare_ksz_map}.
\textbf{\textit{Bottom:}} The 2D power spectra of the patchy kSZ maps from the two simulations. To disentangle the effect from the velocity difference, we also show the spectrum calculated using the AMBER $n_e$ field with the velocity from RadHydro (\textit{dashed purple}). We find good agreement at $\ell \sim 3000$. The difference at $\ell<6000$ mainly results from the electron number density as opposed to the velocity.}
\label{fig:compare_ksz_stats}
\end{figure}

In this section, we present the comparison of the free electron number density, the patchy kSZ 2D projected maps, and the patchy kSZ power spectra between AMBER and RadHydro Sim 1. 

Since our target resolution when running AMBER is $1\,{\rm Mpc}/h$, we first bin down the RadHydro Sim 1 to $64^3$ cells in the $50\,{\rm Mpc}/h$ box. 
Then we re-measure the reionization parameters in the binned Sim 1 and obtain $z_{\rm mid} = 7.85$, $\Delta_z = 4.73$, $A_z = 2.25$.
Note that this is slightly different from the parameter measured in the original resolution shown in Table \ref{tab:scorch} because the smoothing scale is longer.
For parameter calibration and comparisons, we run AMBER at the same resolution ($64^3$), with the same cosmological parameters as RadHydro Sim 1, and we match the reionization parameters measured above.
Because we are using a small box here, we can calculate the maps under the flat sky approximation and do the 2D projection by summing the field along the $z$-axis at a fixed redshift.

\subsection{Free Electron Number Density}

We first examine the evolution of the free electron number density, as it is a crucial component for calculating both the Thomson optical depth and the patchy kSZ signal. 
The fluctuation in free-electron number density will dominate the patchy kSZ signal on small scales.

After getting the reionization redshift field $z_{re}$ following the procedures described in Section \ref{subsec:amber}, we can use it to obtain the electron number density field at all redshifts by 
\begin{equation}
    n_e(\mathbf{x},z) = x_i(\mathbf{x},z) n_b(\mathbf{x},z)(X+\frac{Y}{4}),
\end{equation}
where $n_b$ is the baryon number density, $X$ is the mass fraction of hydrogen, $Y$ is the mass fraction of helium. 
$x_i=n_{\rm e,free}/n_{\rm e,total}$ is the free electron fraction, and is set to be $x_i(\mathbf{x},z)=1$ if $z<z_{re}(\mathbf{x})$ and $x_i(\mathbf{x},z)=0$ if $z>z_{re}(\mathbf{x})$.

In Figure \ref{fig:ne_visual} we show the visualization of a $50 \times 50\times 1\,{\rm Mpc}^3/h^3$ slice of the free electron number density at $z=8$, from the RadHydro Sim 1 and AMBER with matched reionization parameters. 
At $z=8$, about half of the mass in the universe is ionized, and we can see from both simulations that the ionized regions also correspond to the higher-density regions. 
The morphology of ionized regions from the AMBER code is very similar to RadHydro Sim 1. 
However, from the slice we notice that the RadHydro simulation has a larger ionized region in volume.
Given the same mass-weighted ionization fraction, a larger volume-filling factor in RadHydro means that the mean density of the ionized region is smaller, and that the ionizing front propagates further into the IGM.

This is confirmed by the distribution of the gas density in ionized regions shown in Figure \ref{fig:ne_pdf}. 
We can see from the distribution that the ionized regions in AMBER have a peak at higher densities compared to RadHydro, especially during the early stage of reionization ($\bar{x}_{\rm i}<0.5$). 
This happens for two main reasons. 
First, the high-density regions have a high recombination rate and may remain neutral in RadHydro. 
However, this is currently not treated in AMBER, and thus AMBER tends to ionize more high-density regions. 
Second, the RadHydro simulations have episodic star formation and the highest-density collapsed regions do not necessarily produce the highest number of photons. 
These fluctuations show up more when there are small HII regions early on during reionization.
The two processes combined lead to the more tilted PDF of ionized gas density in AMBER.

Then in Figure \ref{fig:ne_power}, we further compare between the AMBER and RadHydro free electron number density power spectra $P_{ee}(k,z)$ at different ionization levels. 
In both simulations, the overall $P_{ee}(k,z)$ gets lower as reionization evolves, because as more mass in low-density regions gets ionized, the ionized electron field becomes a less biased tracer of the matter density field.
At the beginning of reionization, the ionized regions are concentrated in high-density regions around the source galaxies, and therefore $n_e$ has a higher power on large scales compared to the underlying matter density field. 
From the comparison between the two simulations, we see that when reionization just begins ($z\sim 12$), AMBER has larger power on large scales.
As we have already seen in Figure \ref{fig:ne_pdf}, AMBER ionizes more high-density regions during the beginning of reionization. This leads to more bias on large scales at high redshifts.

From $z=9$ onwards, the two simulation matches well with each other, because the bias in AMBER is less prominent as the ionizing fronts propagate further into the low-density regions.
Finally, once reionization is almost over ($z\sim 6$) and all IGM atoms are ionized, the fluctuations in free electrons density follow those of dark matter on large scales.

\subsection{Patchy kSZ}
Next, we compare the 2D projected patchy kSZ temperature maps under flat sky approximation for RadHydro and AMBER.
To get the flat-sky maps, we simply sum the free-electron momentum along the $z$-axis to get the fractional temperature difference in the CMB introduced by the patchy kSZ effect, and then multiply by the CMB temperature $T_{\rm CMB} = 2.725$k to obtain the temperature fluctuation in the 2D plane. 
Here we assume a fixed $z=z_{\rm mid}$ ($x_{\rm HII}=0.5$) since the box size is small enough for ignoring the redshift evolution.

From the projected maps in Figure \ref{fig:compare_ksz_map}, we can see that visually, the AMBER kSZ signal resembles that of RadHydro Sim 1.
The blue and red regions correspond to the signal from the ionized regions along the line-of-sight, while the small white regions are neutral.
Similar to what we have seen in Figure \ref{fig:ne_visual}, because RadHydro has larger ionized regions, we see fewer white pixels in the kSZ map here.

Besides the visualization, in Figure \ref{fig:compare_ksz_stats} we further show the one-point and two-point statistics for the projected kSZ maps. The distributions of the temperature fluctuations shown in the top panel are very similar in the two simulations. 
From the PDF we do see a higher peak around $\Delta T=0$ from AMBER, as we have discussed from the maps in the previous paragraph. 
Other than the difference in the peak, the overall shape and width of the distribution match well. 

The bottom panel shows the dimensionless power spectrum of the flat-sky maps, related to the angular power spectrum $C_\ell$ by $D_\ell=\ell(\ell+1)C_\ell/2\pi$.
We first notice that from both simulations, the power spectrum peaks at $\ell \sim 3500$. 
This scale corresponds to a size of the ionized bubbles of $\sim 11\,{\rm Mpc}/h$ at $z=7.8$. 
Then, comparing the two curves, we see that the spectrum of RadHydro has $\sim 5\%$ higher power on small and large scales, while AMBER produces $\sim 10\%$ higher power at $\ell=3500- 4000$.
There are two sources of the differences: first, RadHydro uses a $P^3M$ N-body simulation to generate the velocity field, while AMBER uses 2LPT and produces less power on small scales compared to N-body \citep[see e.g.][]{Trac2021}; 
second, as was shown in the previous section, there are also differences in the free-electron number density field due to the slightly different morphology of ionized regions.
To disentangle the two effects, we also show the spectrum calculated using the AMBER $n_e$ field with the velocity from RadHydro (shown in dashed purple).
We can see that when we correct for the difference from the velocity fields, the power on small scales ($\ell>6000$) matches perfectly with RadHydro.
This indicates that the small-scale difference is primarily due to the coupling of the $n_e$ fields with the velocity fluctuations.
For $\ell<6000$ multipoles, however, using the RadHydro velocity does not change the kSZ spectrum in AMBER.
This tells us that the $\ell<6000$ arises from the difference in the free electron number density.
Despite the minor differences we just discussed, we still find that overall the kSZ 2D spectrum from AMBER matches well with RadHydro. 


\section{Parameter Space Study}
\label{sec:params}

\begin{figure}[t]
\includegraphics[width=0.5\textwidth]{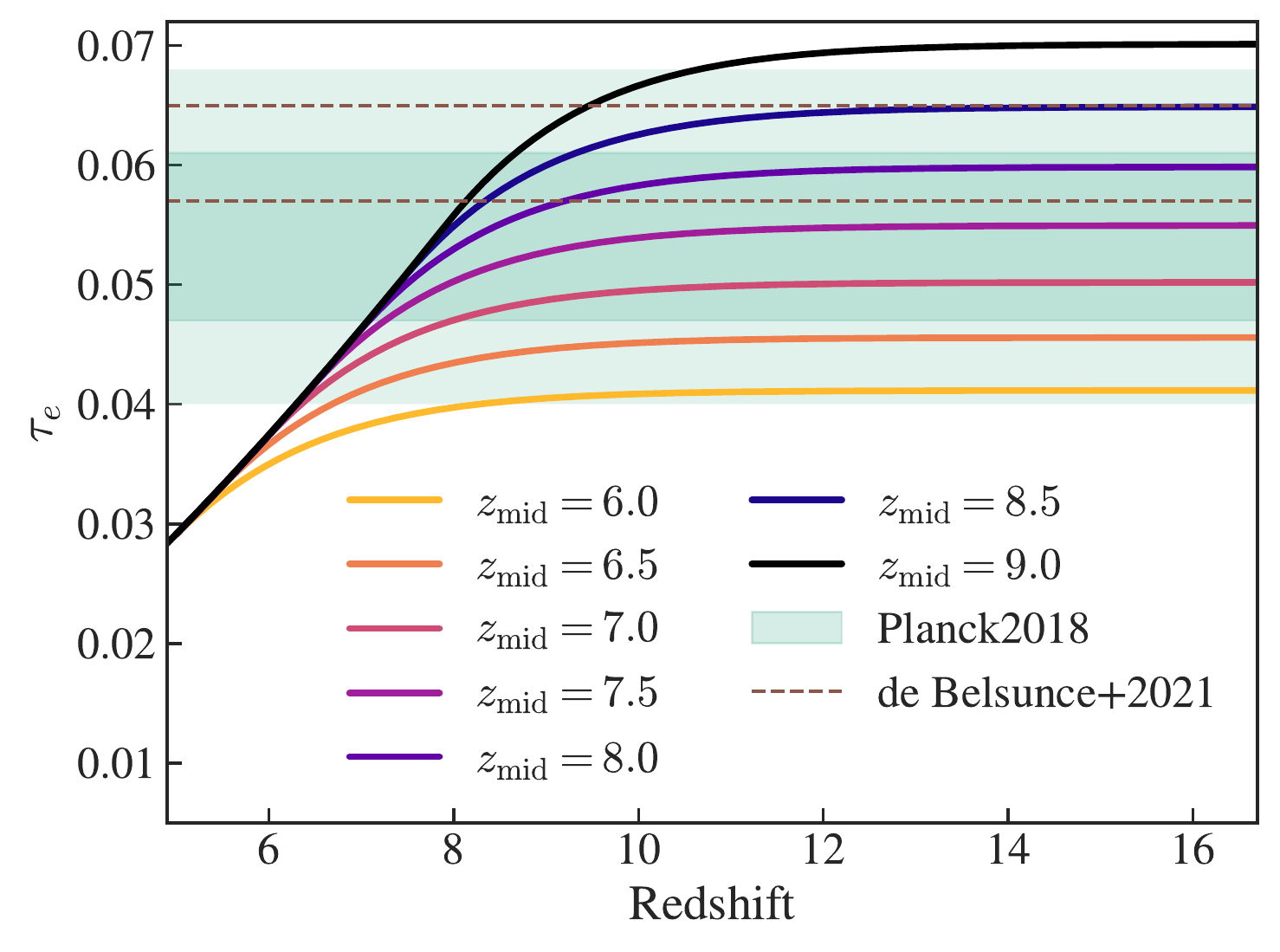}
\caption{The evolution of $\tau_e$ with different values of $z_{\rm mid}$ ranging from $z_{\rm mid}=6.0$ to $z_{\rm mid}=9.0$. 
Out of the values shown, $z_{\rm mid}=9.0$ are mildly inconsistent with the constraint from \citet{Planck2018arXiv180706209P} at the $2\sigma$ level, while the other values are consistent. 
The dashed band shows the re-analysis of the Plank2018 data by \citet{deBelsunce2021MNRAS.507.1072D}, who found larger values of $\tau_e$ and favors higher $z_{\rm mid}$.} 
\label{fig:tau}
\end{figure} 

\begin{figure*}[t]
\centering
\includegraphics[width=\textwidth]{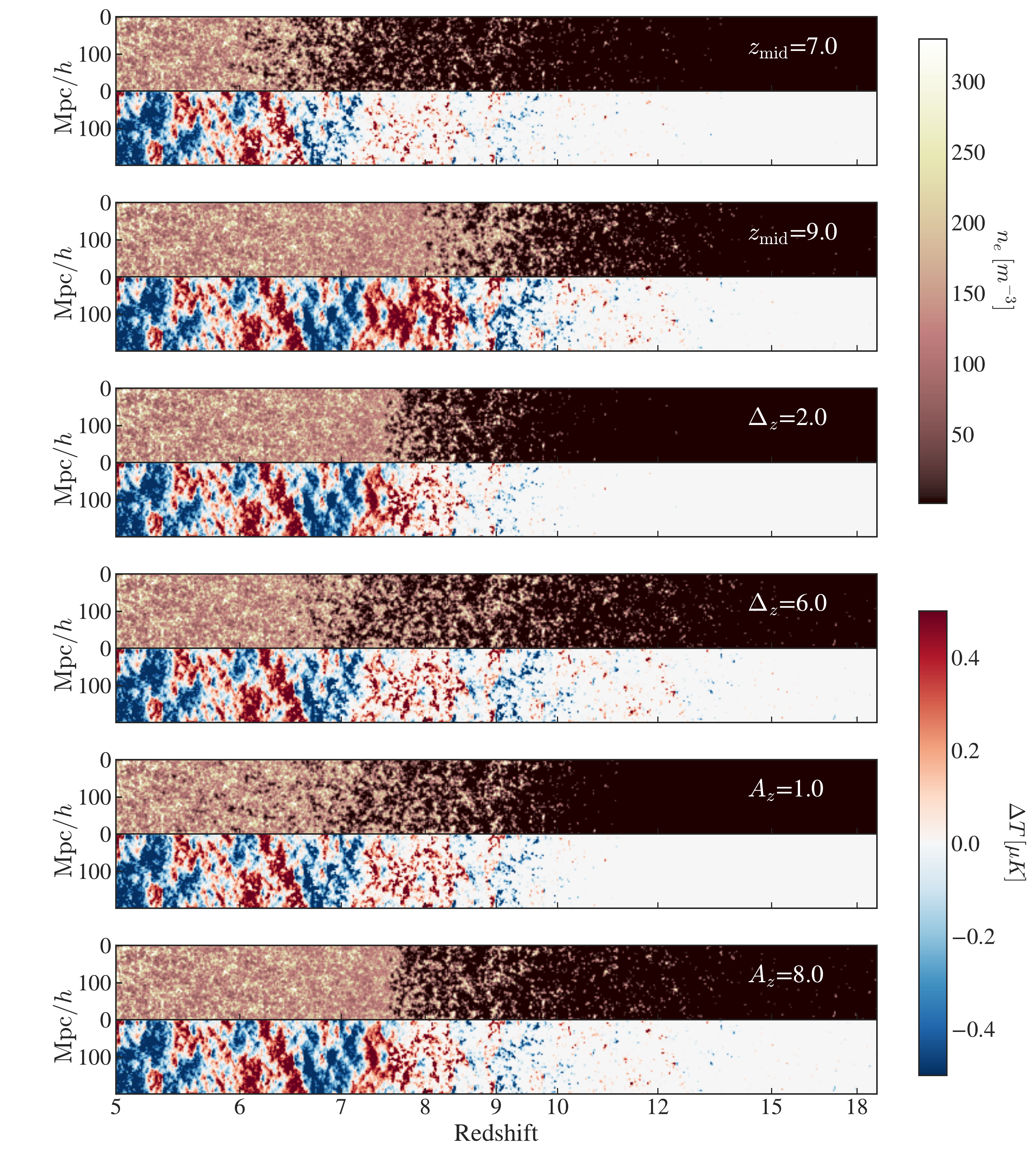}
\caption{Visualization of the redshift evolution of $n_e$ (\textit{dark background}) and $\Delta T_{\rm kSZ}$ (\textit{white background}).
\textbf{\textit{Row 1/2:}} we vary the midpoint of reionzation from $z_{\rm mid}=7.0$ to $z_{\rm mid}=9.0$, while keeping other parameters at their fiducial values.
\textbf{\textit{Row 3/4:}} we vary the duration of reionzation from $\Delta_z=2.0$ to $\Delta_z=6.0$. We can see the large-scale velocity coherence across redshifts. With a longer duration, there is more ionizing bubbles stacked along the line-of-sight.
\textbf{\textit{Row 5/6:}} we vary the asymmetry of reionzation history from $A_z=1.0$ to $A_z=8.0$. With a larger $A_z$, ionizing bubbles begin to form as early as $z=18$, although $z_{\rm mid}$ and $\Delta_z$ is kept fixed.} 
\label{fig:light-cone}
\end{figure*}

\begin{figure*}[t]
\centering
\includegraphics[width=\hsize]{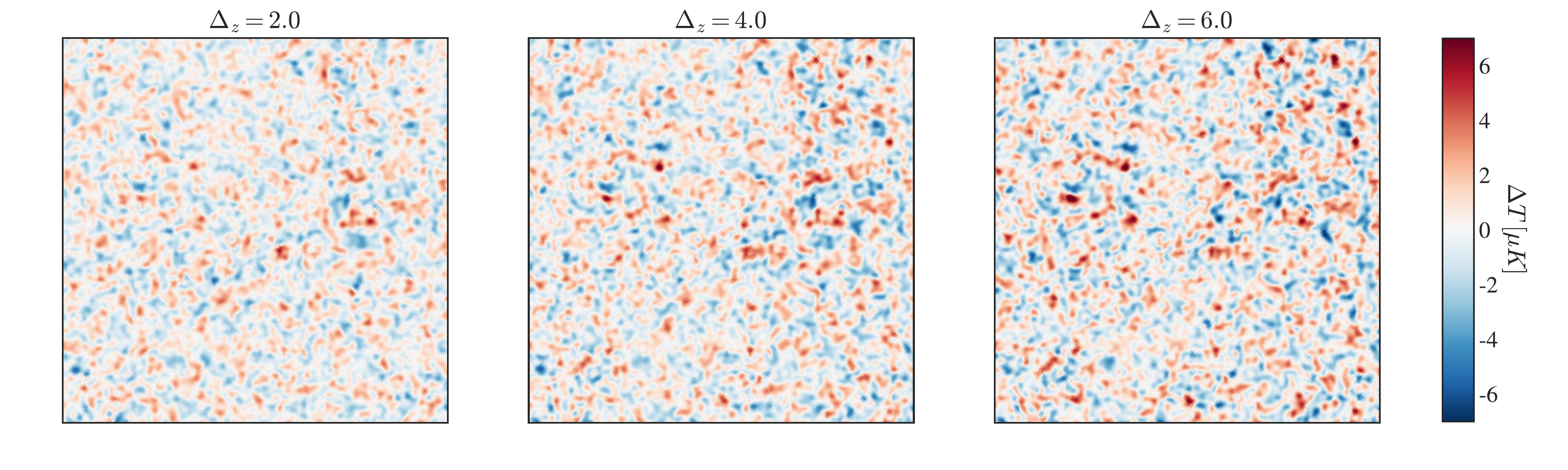}
\caption{$4\times 4$ ${\rm degree}^2$ maps of the kSZ temperature fluctuation for different durations of reionization. This is a filtered map with only the small-scale ($\ell>1000$) modes to show the effect of patchy reionization rather than the large-scale velocity fluctuation.
A longer duration ($\Delta_z=6$, \textit{right}) leads to larger fluctuations on small scales, while the map with a shorter duration appears smoother. 
This is because the small-scale kSZ is sourced by the electron number density fluctuation from patchy reionization. 
These fluctuations are incoherent and accumulate along the line of sight, leading to a larger small-scale inhomogeneity for a longer duration.}
\label{fig:dz_maps}
\end{figure*}

Having calibrated the parameters in AMBER such that the observations at the fiducial parameters match well with the RadHydro simulations, now we vary the parameters around their fiducial values in order to study their effect on the patchy kSZ signal. 
Previous works \citep[e.g.][]{Zahn2007ApJ...654...12Z,Mesinger2012MNRAS.422.1403M,Zahn2012ApJ...756...65Z,Battaglia2013ApJ...776...83B} have shown that the kSZ signal from patchy reionization depends on the midpoint redshift and duration of reionization. 
However, they did not directly parametrize the simulations with these parameters, so it is hard to control the reionization history and directly study its influence on the patchy kSZ signal. 
In this work, we will directly examine the effect of reionization history, parametrized by the midpoint, duration, and asymmetry, on the patchy kSZ effect of reionization.
To generate the kSZ maps and power spectra shown in this section, we run $L_{\rm box}=2\,{\rm Gpc}/h$ simulations on $2048^3$ grids.

\subsection{Redshift Midpoint}
The midpoint redshift $z_{\rm mid}$ is the redshift at which half of the universe is ionized (by mass). 
When other parameters are kept fixed, a higher $z_{\rm mid}$ means that the whole reionization process is pushed to an earlier time when the universe has a higher energy density. 
We note that because we currently generate the reionization redshift field by abundance matching at $z_{\rm mid}$, a change in $z_{\rm mid}$ can also affect the relative order of ionization of the cells (i.e. we do not preserve the exact same ionization morphology by fixing all other parameters). 
However, we expect such an effect to be small, because the large halos at $z=9$ should also correspond to large halos at $z=7$.
Therefore, we expect the main effect from varying $z_{\rm mid}$ to be the amplitude of the kSZ spectra:  the overall amplitude should be larger with a higher $z_{\rm mid}$, because there would be a higher electron density when we integrate along the line of sight. 

In the top two panels of Figure \ref{fig:light-cone}, we visualize the free electron number density $n_e$ and the kSZ temperature change $\Delta T_{\rm kSZ}$ for a relatively early ($z_{\rm mid}=9$) and late ($z_{\rm mid}=7$) reionization.
In this figure, the horizontal axis is the line-of-sight, and in the direction perpendicular to the page we plot a $1\,{\rm Mpc}/h$ slice from the 3D simulation box. 
From the $n_e$ plots, we can see ionized bubbles form around the first galaxies at the beginning of reionization. 
These bubbles continue to grow in size and finally overlap and merge, leading to a fully ionized universe.
In the left panel of Figure \ref{fig:ksz_3_reion_params}, we can see that increasing the midpoint redshift of reionization increases the amplitude of the kSZ spectra without changing the shape.
Physically, the scenario that corresponds to an earlier $z_{\rm mid}$ could be a higher escape fraction, the ionization dominated by smaller sources, or a combination of multiple effects. 

We note that there exist degeneracies between $z_{\rm mid}$ and other reionization parameters (most noticeably the duration of reionization), in terms of their effects on the patchy kSZ power spectrum (we will discuss more about it later). 
Such degeneracy is often broken by constraints from the Thompson optical depth $\tau_e$ from the low-$\ell$ $EE$ polarization \citep[e.g.][]{Ferraro2018}.
The value of $\tau_e$ is mainly affected by the redshift of reionization without being sensitive to other reionization parameters \citep[e.g.][]{Battaglia2013ApJ...776...83B}. 
In Figure \ref{fig:tau}, we show the evolution of $\tau_e$ with different values of $z_{\rm mid}$ ranging from $z_{\rm mid}=6.0$ to $z_{\rm mid}=9.0$. 
Out of the values shown, $z_{\rm mid}=9.0$ is mildly inconsistent with the constraint from \citet{Planck2018arXiv180706209P}, while the other values are all within the $2\sigma$ range.
However, we note that recent re-analysis of the Planck2018 data by \citet{Pagano2020A&A...635A..99P} and \citet{deBelsunce2021MNRAS.507.1072D} (shown as dashed lines) found larger values of $\tau_e$ and favors higher $z_{\rm mid}$. 
In either case, the constraint we get from $\tau_e$ on $z_{\rm mid}$ is tighter compared to the constraint from purely patchy kSZ.

\subsection{Duration}
\label{sec:duration}

In the previous section, we have hinted at the degeneracy between $z_{\rm mid}$ and duration of reionization.
Now we will turn to the effect of the duration on the patchy kSZ temperature fluctuations.
Out of all the reionization parameters, $\Delta_z$ has the strongest effect on the amplitude of the spectra. 
For this reason, observations of the kSZ amplitude have been used to constrain the reionization duration for the past decade.
The ACT and SPT-SZ surveys have published upper limits on the kSZ power \citep{Addison2013,Dunkley2013JCAP...07..025D,George2015ApJ...799..177G}, with a $95 \%$ CL upper limit on the patchy kSZ power being $D^{\rm pkSZ}_{\ell=3000} < 3.3\mu K^2$ from the 2500 ${\rm degree}^2$ SPT-SZ survey.
By combining SPT results with large-scale CMB polarization measurements, \cite{Zahn2012ApJ...756...65Z} constrains the amplitude of the patchy kSZ by setting an upper limit $D^{\rm pkSZ}_{\ell=3000} \leq 2.1 \mu K^2$  (95\% CL).
The most recent observational constraints come from the SPTPol survey, where \cite{Reichardt2021ApJ...908..199R} constrained the patchy kSZ amplitude to $D^{\rm pkSZ}_{\ell=3000} = 1.1^{+1.0}_{-0.7} \mu K^2$ using the models of homogeneous signal given in \cite{Shaw2012}. 
Using the fitting template provided in \citet{Battaglia2013ApJ...776...83B}, they find the $95\%$ CL upper limits on the duration of reionization to be $\Delta_{z,50}<5.4$ (6.9/4.3 when considering a $25\%$ uncertainty in the homogeneous spectrum), and a $68\%$ confidence interval of $\Delta_{z,50}=1.1^{+1.6}_{-0.7}$.
Using a more recent semi-numerical model, \citet{Choudhury2021} placed a tighter constraint on the duration to be $\Delta_{z,50}<2.9$ at $99\%$ CL.

Before diving into the kSZ angular power spectra prediction with AMBER, we start by visualizing the patchy kSZ temperature fluctuations with different durations of reionization. 
In the middle panels of Figure \ref{fig:light-cone}, we show the light-cone projections of the free electron number density $n_e$ and the kSZ temperature change $\Delta T_{\rm kSZ}$ assuming a quick reionization ($\Delta z=2$, third row) and a slow reionization ($\Delta z=6$, fourth row). 
From the comparison between the two durations, we see that under the $\Delta z=6$ scenario, reionization starts earlier, and there are more ionized bubbles along the line-of-sight. 
The motion of these ionized bubbles relative to the background CMB would result in the observed kSZ fluctuation in the CMB we see from the bottom panels.
In the bottom panels, we can clearly see the large-scale velocity fluctuations. 
These fluctuations dominate $\Delta T_{\rm kSZ}$ on large scales, while the $n_e$ field fluctuation dominate on small scales.

In Figure \ref{fig:dz_maps}, we show $4\times 4$ ${\rm degree}^2$ patches of the $\Delta T_{\rm kSZ}$ maps from the $2\,{\rm Gpc}/h$ simulations with different durations. 
These maps are made by ray-tracing through past light cones during reionization ($5 \leq z \leq 30$), and generated using \texttt{HEALPix} \citep[][]{Gorski2005} with $N_{\rm side}=4 \times N_{\rm mesh}$.
The maps shown here are processed with a high-pass filter, where we only keep the spherical harmonics with $\ell>1000$.
This is because we are more interested in the small-scale features which is a direct result of patchy reionization. 
From the maps, we can see that a longer duration leads to larger fluctuations on small scales, because the small-scale kSZ is sourced by the electron number density fluctuation from patchy reionization. 
These fluctuations are incoherent and accumulate along the line of sight, leading to a larger small-scale inhomogeneity for a longer duration.

\begin{figure*}[t]
\centering
\includegraphics[width=1.\textwidth]{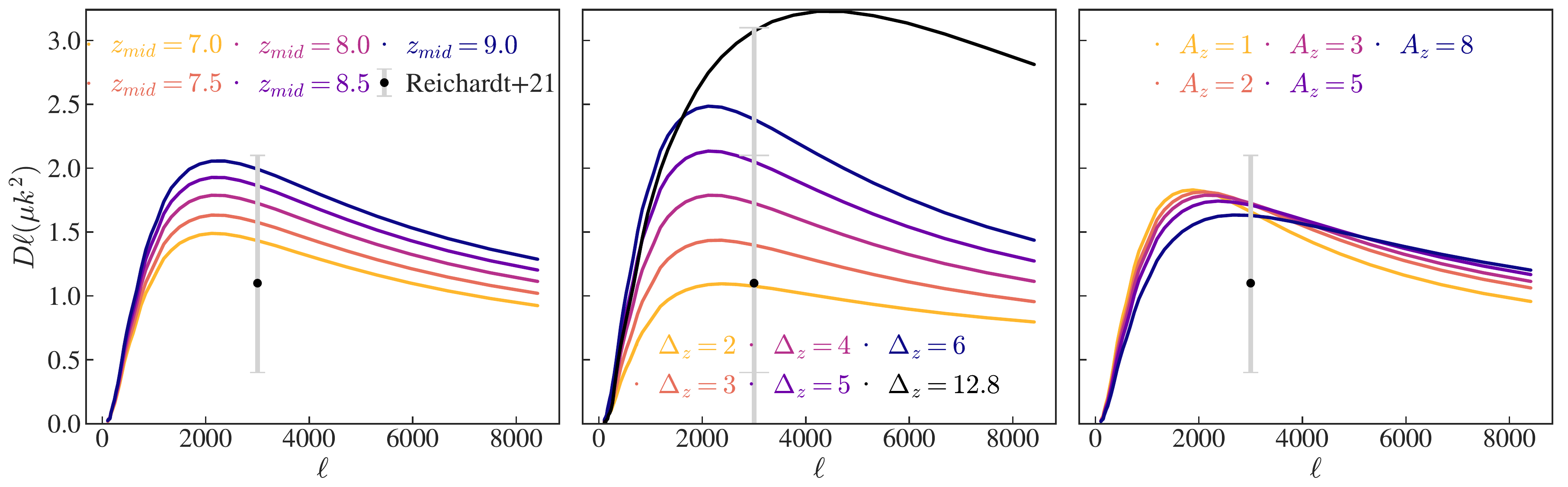}
\caption{Patchy kSZ angular power spectrum for different reionization history parameters. The fiducial parameters are $[z_{\rm mid},\Delta_z,A_z]=[8.0,4.0,3.0]$. The smaller error bars show the $1\sigma$ confidence interval of $D^{\rm kSZ}_{\ell=3000}$ from \citet{Reichardt2021ApJ...908..199R}, and the larger error bar in the middle panel shows the $2\sigma$ confidence interval. \textbf{\textit{Left:}} The overall amplitude of the kSZ spectrum increases as we shift the midpoint redshift of reionization earlier, but all variations are consistent with the $1\sigma$ interval. \textbf{\textit{Middle:}} Increasing the duration of reionization also increases the overall power of the kSZ spectrum, and it affects the kSZ amplitude most significantly. The black line shows the maximum $\Delta_z$ that produces $D^{\rm kSZ}_{\ell=3000}$ within the $2\sigma$ interval, where we let $z_{\rm mid}=6.5$, $A_z=8$ (to ensure reionization ended before $z=5.5$), and $\lambda_{\rm mfp}=1.0\,{\rm Mpc}/h$. \textbf{\textit{Right:}} The asymmetry parameter $A_z$ does not have a big impact on the kSZ spectrum comparing with the midpoint and duration, but we can see that increasing the asymmetry (meaning that the beginning of reionization is longer) results in flatter slope of the kSZ spectrum.}
\label{fig:ksz_3_reion_params}
\end{figure*}

Finally, we show the change in the angular power spectra with respect to the duration in the second panel of Figure \ref{fig:ksz_3_reion_params}. 
Compared with the first panel on $z_{\rm mid}$, we see a strong degeneracy of the two parameters as expected, but the spectra are more sensitive to duration than the midpoint redshift.
Here we also show the $1\sigma$ and $2\sigma$ confidence intervals derived in \citet{Reichardt2021ApJ...908..199R}, in order to show the extent of variation with the $\Delta z$ parameter within the confidence intervals. 
For the colored lines, we keep all other parameters at their fiducial values, and we can see that durations of $\Delta_z<5.1$ yield results consistent with the $1\sigma$ constraints from \cite{Reichardt2021ApJ...908..199R}.
This translates to $\Delta_{z,50}<2.0$ under the definition of duration in \cite{Battaglia2013ApJ...776...81B}, assuming a mildly asymmetric reionization history at $A_z=3$ (see Figure \ref{fig:d50_d90} for the conversion). 
The limit agrees with the recent picture from a variety of observations arguing that reionization happened quickly.

Then we explore the maximum $\Delta_z$ that produces a $D^{\rm pkSZ}_{\ell=3000}$ consistent with the $2\sigma$ constraint from \citet{Reichardt2021ApJ...908..199R}.
In order to do this, we also vary the other two parameters, $z_{\rm mid}$ and $\lambda_{\rm mfp}$, that affect $D^{\rm pkSZ}_{\ell=3000}$ most.
Since we know that lower values of $z_{\rm mid}$ and $\lambda_{\rm mfp}$ decreases $D^{\rm pkSZ}_{\ell=3000}$, we want $z_{\rm mid}$ and $\lambda_{\rm mfp}$ to be low in order to allow for a longer duration.
We set $z_{\rm mid}=6.5$, a value slightly below the $1\sigma$ interval from the $\tau_e$ prediction from \citet{Planck2018arXiv180706209P}. 
We also set $\lambda_{\rm mfp}=1\,{\rm Mpc}/h$, and $A_z=8.0$ in order to ensure an end of reionization by $z=5.5$.
After minimizing $D^{\rm pkSZ}_{\ell=3000}$ with other parameters in consistency with other observation channels, we find that duration of $\Delta_z=12.8$ reaches the top of the $2\sigma$ interval.
This converts to $\Delta_{z,50}<3.5$ under the assumption that $A_z=8.0$. 

We emphasize, however, that this should not be interpreted as a strict $2\sigma$ constraint of $\Delta_z <12.8$ (or $\Delta_{z,50}<3.5$), because here we do not systematically vary other reionization parameters. 
A full parameter space study is needed in order to derive a constraint on the duration from the simulation data and the observation data. 
We also note that our $D^{\rm pkSZ}_{\ell}$ are integrated from $z=5$ to $z=30$, while the confidence interval derived in \citet{Reichardt2021ApJ...908..199R} is derived using an end of reionization redshift of $z=5.5$. 
Converting to an end of reionization redshift of $5.0$ may move the interval up by $\sim 5-10\%$. Finally, the uncertainty within the homogeneous kSZ spectra can cause $\sim 25\%$ fluctuations in the patchy kSZ estimation, according to \citet{Reichardt2021ApJ...908..199R}.

\subsection{Asymmetry}

\begin{figure*}[t]
\includegraphics[width=0.48\hsize]{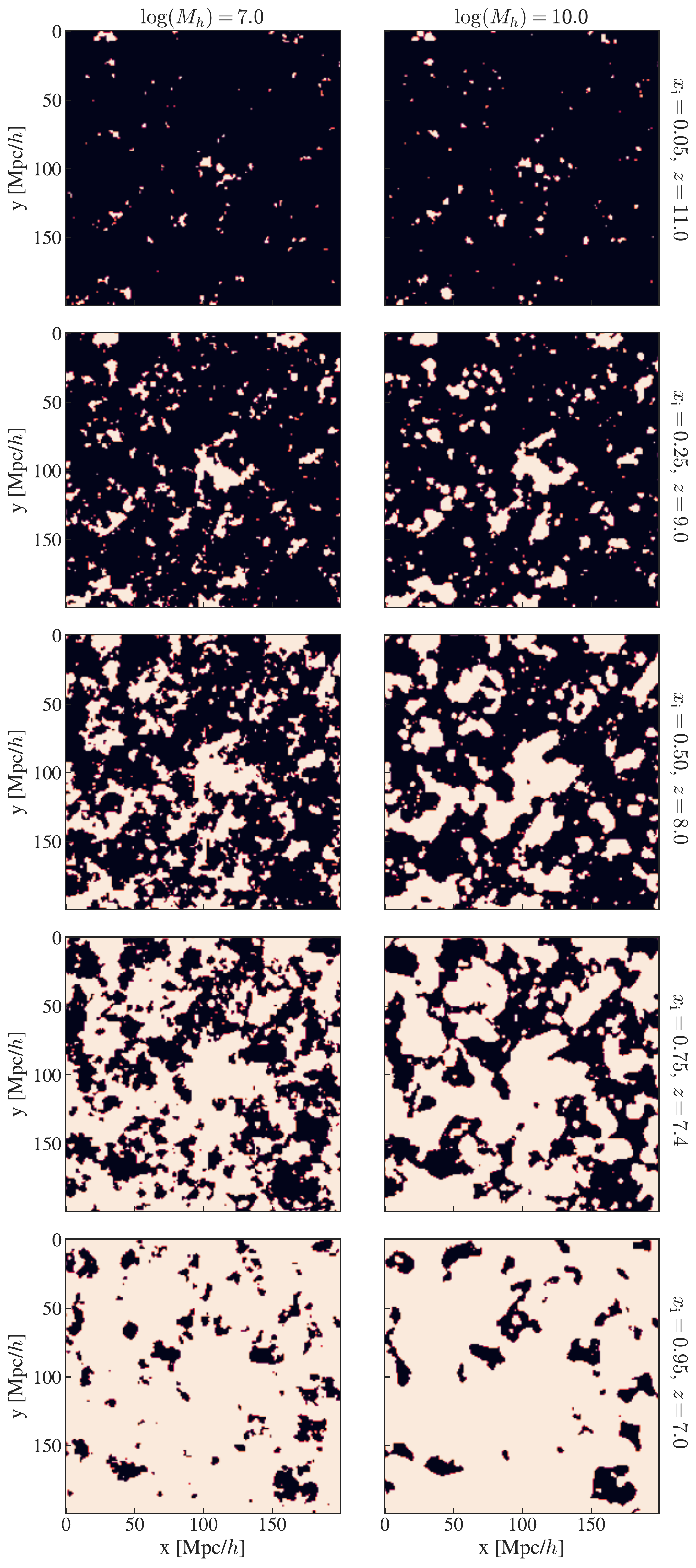}
\includegraphics[width=0.48\hsize]{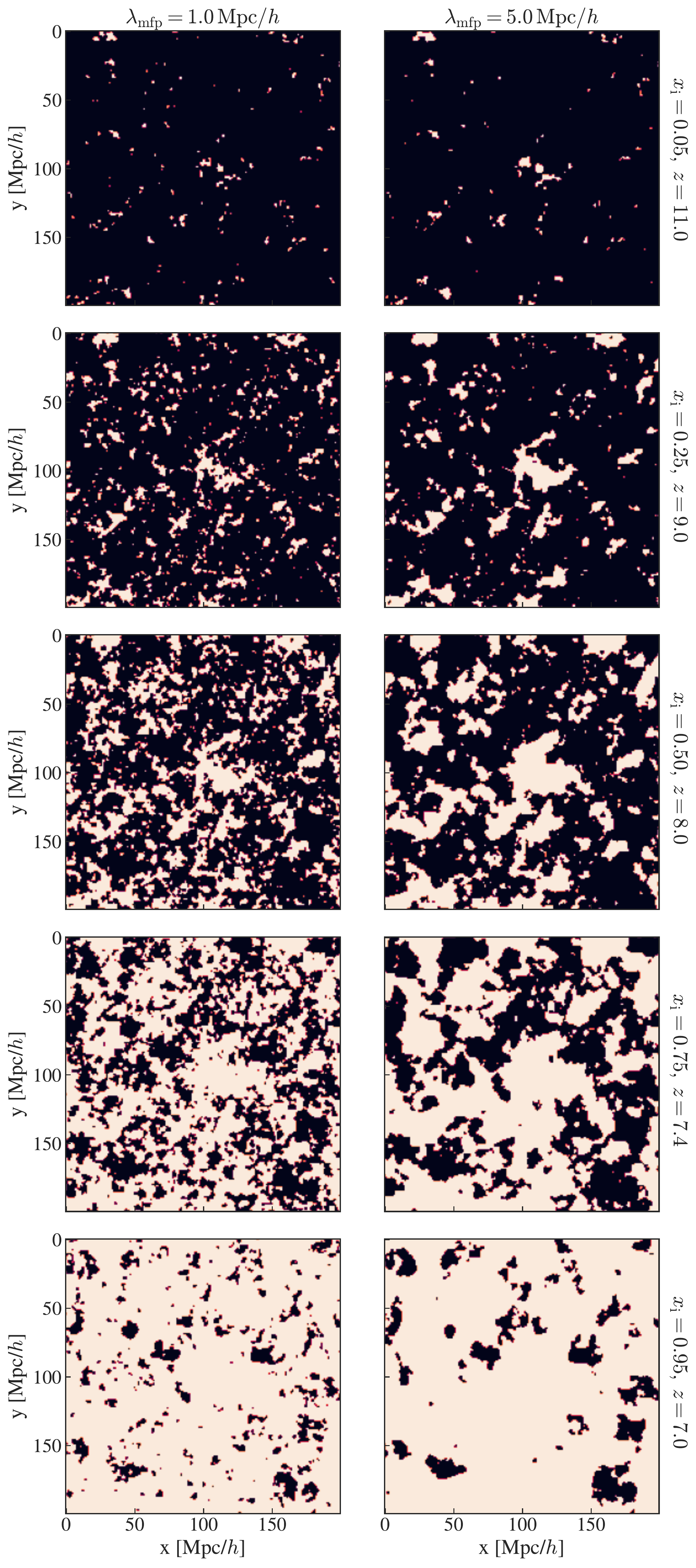}
\caption{Ionization fraction field across different (mass-weighted) ionization levels, with white regions marking the ionized bubbles. From top to bottom with show a $(200\,{\rm Mpc}/h)^2 \times 1\,{\rm Mpc}/h$ slice at $\bar{x}_i=5\%$, $25\%$, $50\%$, $75\%$, and $95\%$ , respectively. \textbf{\textit{Left:}} ionized regions with a minimum halo mass of $10^7\,M_\odot$ (\textit{first column}) and $10^{10}\,M_\odot$ (\textit{second column}). With very large minimum halo mass for ionizing galaxies, the ionized bubbles are smoother and more clustered on large scales. However, the overall morphology are not drastically different from when $M_{\rm h}=10^7 M_\odot$.
\textbf{\textit{Right:} }ionized regions with $\lambda_{\rm mfp} = 1.0\,{\rm Mpc}/h$ (\textit{third column}) and $\lambda_{\rm mfp} = 5.0\,{\rm Mpc}/h$ (\textit{fourth column}). 
We see that in AMBER, with a fixed reionization history, $\lambda_{\rm mfp}$ has a stronger effect on the ionization morphology than the minimum halo mass $M_h$.
With a smaller $\lambda_{\rm mfp}$, the typical sizes of ionized regions are significantly smaller than with a larger $\lambda_{\rm mfp}$.}
\label{fig:ion_fields}
\end{figure*}

\label{sec:asy}

\begin{figure*}[t]
\centering
\includegraphics[width=0.98\hsize]{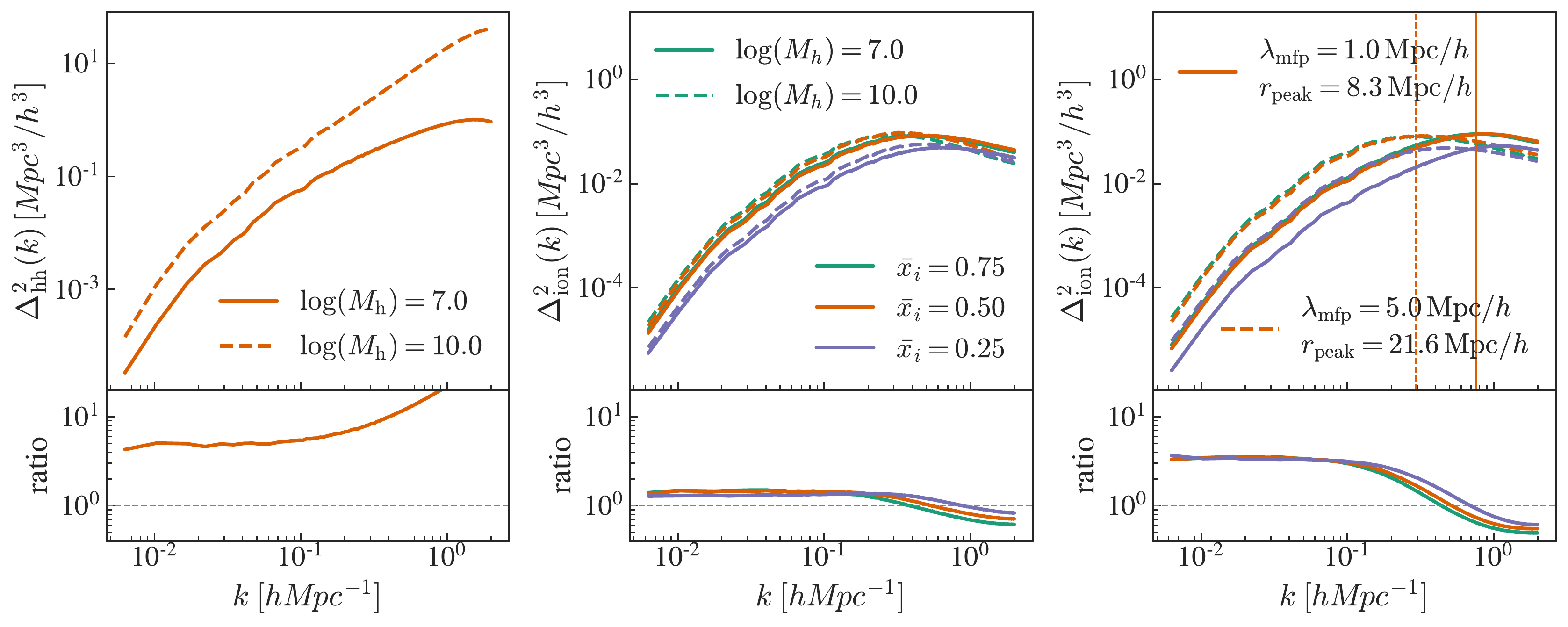}
\caption{Dimensionless power spectra of the AMBER halo density fields (\textit{left}) and ionization fraction fields (\textit{middle and right}) with different minimum halo mass and photon mean free path. For the ionization fraction power spectra we show the spectra at three global ionization levels ($\bar{x}_i=0.25,0.50,0.75$, corresponding to the \textit{purple, orange} and \textit{green} lines, respectively).
\textbf{\textit{Left:}} dimensionless halo density power spectra for $M_h=10^7\,M_\odot$ (\textit{solid}) and $M_h=10^{10}\,M_\odot$ (\textit{dashed}).
\textbf{\textit{Middle:}} $\Delta^2_{\rm ion}(k)$ for $M_h=10^7\,M_\odot$ (\textit{solid}) and $M_h=10^{10}\,M_\odot$ (\textit{dashed}).
\textbf{\textit{Right:}} $\Delta^2_{\rm ion}(k)$ for $\lambda_{\rm mfp}=1.0\,{\rm Mpc}/h$ (\textit{solid}) and $\lambda_{\rm mfp}=5.0\,{\rm Mpc}/h$ (\textit{dashed}).
The bottom panels show the ratio of the power spectra, with the ratio being $P_{\rm ion,log(M)=10.0}/P_{\rm ion,log(M)=7.0}$ in the left/middle panels and $P_{\rm ion,\lambda=5}/P_{\rm ion,\lambda=1}$ on the right.}
\label{fig:ion_pk}
\end{figure*}

The asymmetry parameter $A_z$ characterizes the relative length of the beginning and end of reionization.
\citet{Park2013ApJ...769...93P} showed that the model adopted in \citet{Battaglia2013ApJ...776...81B} failed to account for an asymmetric reionization history, and cannot be used to provide universal modeling of the kSZ spectrum.
In AMBER, the asymmetry parameter $A_z$ allows us to have more control over the overall shape of the reionization history and reduce the modeling bias.
When we set the asymmetry large, we are enforcing an earlier but slower beginning, and an earlier and abrupt end of reionization.
When $A_z \sim 1$, we get a symmetric reionization history where the beginning and end have an equal length.
In the bottom panels of Figure \ref{fig:light-cone}, we show the redshift evolution of $n_e$ and $\Delta T_{\rm kSZ}$ with a symmetric reionization and a highly asymmetric reionization.
We can see that even when the duration is kept fixed, for the asymmetric reionization scenario, the ionizing bubbles begin to form at a much earlier redshift.

The right panel of Figure \ref{fig:ksz_3_reion_params} shows the change in the kSZ spectrum when we only change the asymmetry of the reionization history. 
Compared with the other two reionization history parameters, the dependence of the kSZ spectra on $A_z$ is weak. 
We notice that at the $\ell=3000$ scale where the observation data lies, there is almost complete degeneracy between the different asymmetric reionization cases (but note that the cross-over scale may be different for other fiducial parameters.). 
Yet there is a noticeable change in the slope of the kSZ spectrum: large asymmetry would decrease the power on large scales and introduce slightly more power on small scales. 
This indicates that to further constrain the early and end phase of reionization in addition to the overall length, we will need more data at multipoles other than $\ell=3000$.

\subsection{Minimum Halo Mass}

\begin{figure*}[t]
\centering
\includegraphics[width=0.87\textwidth]{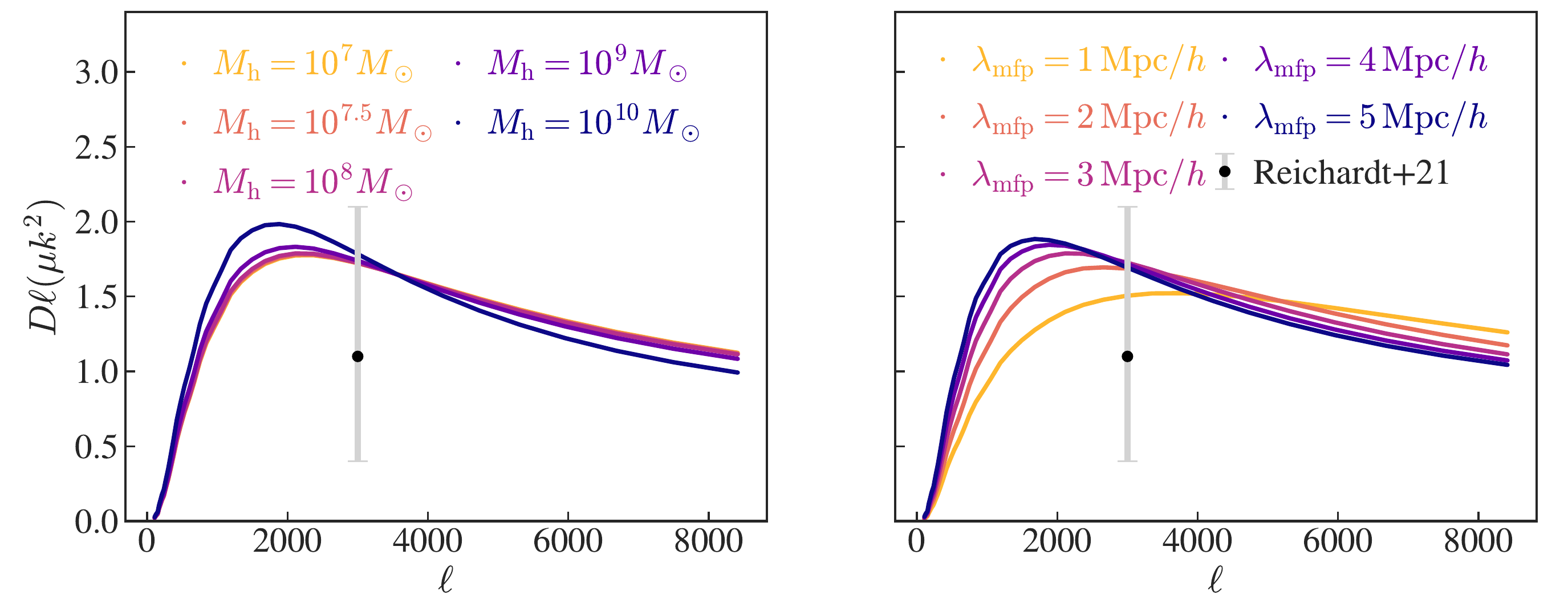}
\caption{\textbf{\textit{Left:}} $\lambda_{\rm mfp}$ is correlated with the average bubble size of ionized bubbles. Increasing the mean free path shifts the peak towards higher multipoles in the kSZ spectrum. 
\textbf{\textit{Right:}} Out of all the parameters, $M_h$ has the least effect on the patchy kSZ signal.
Only when we limit the sources to be above $10^{9}\, M_\odot$ can we see some suppression on the small scale power.}
\label{fig:ksz_2_reion_params}
\end{figure*}

The minimum halo mass parameter, $M_{h}$, is a lower mass limit of halos that host ionizing sources in the simulation. 
Usually in simulations where we do not keep the reionization history and the photon mean-free path fixed, a smaller $M_{\rm h}$ can lead to earlier reionization because the ionizing sources are more abundant at high redshifts. 
However, this effect can be counter-balanced if the ionizing photon budget is small (e.g. the escape fraction is low).
By directly controlling the reionization history, however, we do not have to explicitly account for the degeneracy between the sources and sinks.

In the left two columns of Figure \ref{fig:ion_fields}, we show the evolution of the ionization fraction field assuming two extreme values in our parameter study, $M_h=10^7 M_\odot$ and $M_h=10^{10} M_\odot$, while fixing all other parameters at their fiducial values.
Because the reionization history is fixed, at each redshift the global ionization fractions are the same.
We can see that the change in $M_h$ affects the morphology of ionized regions, but not very significantly. 
Before the Universe gets half ionized, larger $M_h$ leads to more large-scale clustering of ionized regions around heavier sources, and there are fewer small ionized regions.
In AMBER, when we change $M_{h}$ from $10^{10} M_\odot$ to $10^7 M_\odot$, the smaller halos at $z=z_{\rm mid}$ will no longer be treated as sources. 
However, such effect is small if we only care about the rank-ordering of the radiation field: even when the smaller sources are turned on, they will have less radiation compared to large sources, and therefore have a lower priority in reionization compared to larger sources. 
On the other hand, compared with the non-source regions, the small sources are likely nearer to the large sources than under-dense regions are because of the clustering in structure. 
Hence, even when smaller sources are not turned on, they are still likely ionized earlier than the under-dense regions.
In this way, as long as the reionization history is fixed, the $M_h$ value will have a minor effect on the reionization morphology, as the relative order of reionization is mostly preserved.

A more quantitative characterization of the difference in clustering is shown in the left and middle panels of Figure \ref{fig:ion_pk}, where we plot the power spectrum of the halo density field and ionization fraction field at $\bar{x}_i=25\%,50\%$ and $75\%$ for the two $M_h$ values. 
From the halo density spectra shown on the left, we see that there is a constant rise in halo bias by a factor of $\sim 3$ on large scales, while on scales above $k=1\,{\rm Mpc}^{-1}h$ the bias increases to $>10$.
The bias contrast in the ionization fraction field, however, is not as significant. On large scales ($k<0.2\,{\rm Mpc}^{-1}h$), there is a constant increase of power in the $M_h=10^{10} M_\odot$ field, but only by $<5\%$.
On smaller scales, the power of the $M_h=10^{10} M_\odot$ ionization fraction field falls compared with the $M_h=10^{7} M_\odot$ field by $\sim 30\%$ near the end of reionization.
 
Now that we understand the effect of $M_h$ on the ionization morphology, in the right panel of Figure \ref{fig:ksz_2_reion_params}, we show the dependency of $D^{\rm pkSZ}_{\ell}$ on $M_h$, while keeping the reionization history fixed.
We can see that out of all the parameters, $M_h$ has the least effect on the patchy kSZ signal, as should be expected from the small changes in the ionization fraction power spectrum.
Only when we limit the sources to be above $10^{9}\, M_\odot$ can we see a slight increase in the angular power at $\ell\sim 2000$ and suppression of the smaller scale power.
This means that a very large $M_h$ can still affect the morphology of ionized regions, even if we fix the reionization history.
While for smaller $M_h$ values, we do not see an effect on the kSZ spectra when we change $M_h$.
Note that this is not generally true if we do not fix the reionization history, because $M_h$ can affect the timing of reionization and thus the kSZ power.

Figure \ref{fig:ksz_2_reion_params} includes a wide range of $M_h$ from $10^7 M_\odot$ to $10^{10} M_\odot$. However, the change in $D^{\rm pkSZ}_{\ell}$ is at most $0.15 \mu K^2$ at the higher multipoles. 
This is in contrast to the result shown in \citet{Paul2021}, who predicts a $30 \sim 60\%$ increase in $D^{\rm pkSZ}_{\ell=3000}$ with a fixed reionization history. 
This is likely due to the different assumptions we made in order to keep the reionization history fully controlled (in our case the abundance-matching scheme, and in their case, a manually-set ionizing efficiency at each time step). 
In the context of our model, we cannot gain many constraints from $D^{\rm pkSZ}_{\ell}$ directly on the ionizing halos. 
We will need to infer such constraints with extra assumptions on astrophysical parameters that link the reionization history and $M_h$.

\begin{figure*}[t]
\centering
\includegraphics[width=0.83\textwidth]{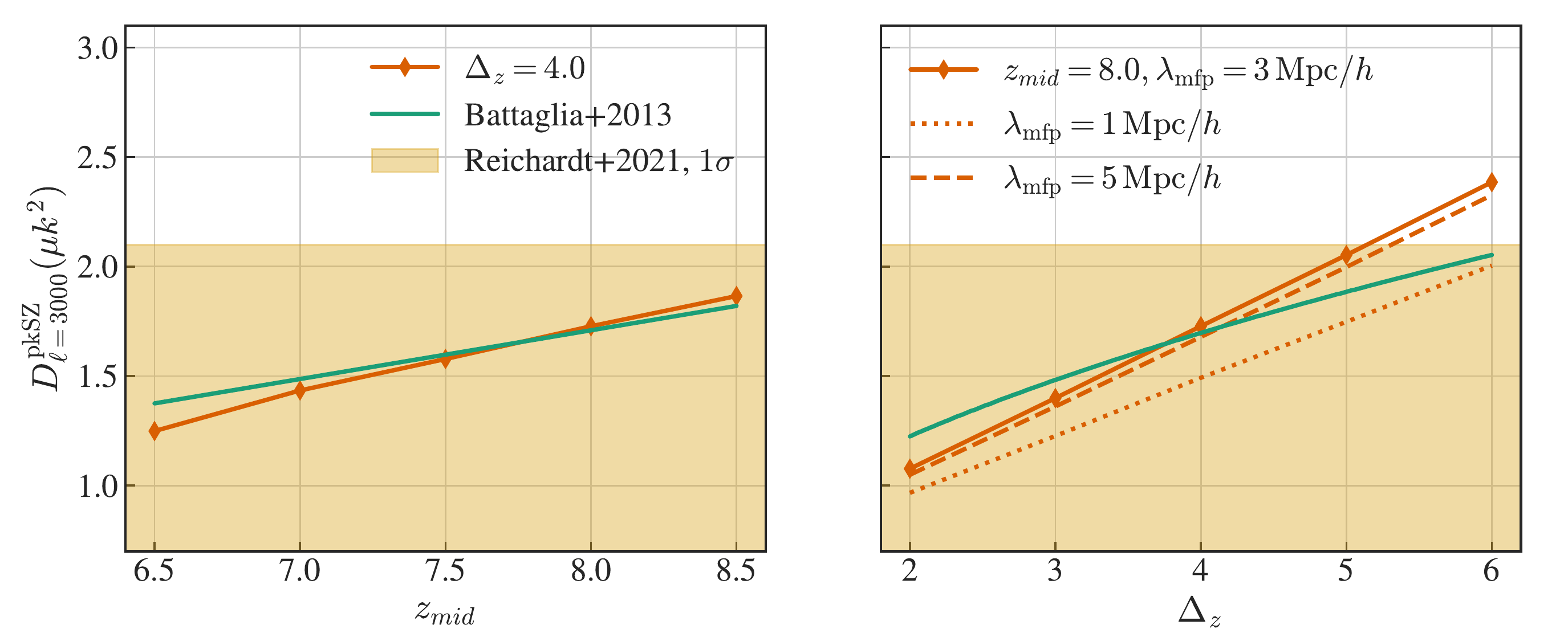} 
\caption{Relationship between the amplitude of the kSZ angular power spectra $D^{\rm pkSZ}_{\ell}$ at $\ell=3000$ and the redshift and duration of reionization. The yellow regions are the $1\sigma$ constraint from \citet{Reichardt2021ApJ...908..199R}. \textbf{\textit{Left:}} with a fixed duration $\Delta_z=4.0$, AMBER produces $D^{\rm pkSZ}_{\ell=3000}$ that scales almost linearly with the midpoint redshift of reioziation (\textit{orange}). Compared with the scaling relation fitted in \citet{Battaglia2013ApJ...776...83B} (\textit{green}), we have a slightly steeper slope. \textbf{\textit{Right:}} when we fix $z_{\rm mid}=8.0$, $D^{\rm pkSZ}_{\ell=3000}$ also scales linearly with the duration of reionization. 
Compared with \citet{Battaglia2013ApJ...776...83B} who found a power-law dependence of $\sim 0.47$, we find a steeper dependence of $D^{\rm pkSZ}_{\ell=3000}$ on $\Delta_z$. We also show the relation at $\lambda_{\rm mfp}=1\,{\rm Mpc}/h$ (\textit{dotted orange}) and $\lambda_{\rm mfp}=5\,{\rm Mpc}/h$ (\textit{dashed orange}), in order to demonstrate the dependence of the scaling relation on $\lambda_{\rm mfp}$.}
\label{fig:scale_dz}
\end{figure*}

\subsection{Radiation Mean Free Path}
Finally, we study the mean free path parameter $\lambda_{\text{mfp}}$ that controls on average how far ionizing photons travel in the IGM before being absorbed. 
Many previous works have empirically related the angular scale at which the patchy kSZ power spectrum reaches its maximum $\ell_{\rm max}$ to the typical size of ionized regions during reionization \citep[e.g.][]{McQuinn2005,Iliev2007ApJ...660..933I,Mesinger2012MNRAS.422.1403M,Gorce2020A&A...640A..90G}.
Under the scenario described in Section \ref{sec:duration}, larger bubbles result in a larger mean free path, as photons travel through the ionized region without being absorbed. 
Therefore, the photon mean free path is strongly correlated with the average bubble size during reionization. 
The ionized bubble size determines the peak of the patchy kSZ spectrum, as we would expect the kSZ spectrum to attain the most power on the scale of the size of these bubbles. 

In AMBER, \textit{the mean free path parameter $\lambda_{\rm mfp}$ is not equivalent to the physical mean free path of photons measured in the IGM.}
Firstly, our $\lambda_{\rm mfp}$ is defined at the midpoint of reionization, as opposed to the usual definition at the end of reionization.
Secondly, $\lambda_{\rm mfp}$ does not directly control the size of the ionized regions at a fixed redshift. 
The sizes of ionized regions depend on the relative radiation intensity as well as the ionization fraction at a specific redshift. 
Hence, we can imagine that even with $\lambda_{\rm mfp}=3\,{\rm Mpc}/h$, the sizes of ionized regions will be much smaller than that at $x_{\rm HII}=0.05$.
Finally, even though the $\lambda_{\rm mfp}$ is a global parameter, it does not mean that the ionized regions all have fixed sizes. 
Large halos will still have larger ionized regions around them, because the photon budget of a cell is affected by the density in addition to the mean free path parameter. 

Similar to the previous section, we begin by visualizing the evolution of ionized regions throughout the EoR with various $\lambda_{\rm mfp}$.
On the right two columns of Figure \ref{fig:ion_fields}, we show the ionization fraction fields with $\lambda_{\rm mfp}=1\,{\rm Mpc}/h$ and $\lambda_{\rm mfp}=5\,{\rm Mpc}/h$.
Compared to $M_h$, we see a larger contrast in the ionization morphology when varying $\lambda_{\rm mfp}$: there are more numerous and smaller ionized bubbles in the $\lambda_{\rm mfp}=1\,{\rm Mpc}/h$ run than the $\lambda_{\rm mfp}=5\,{\rm Mpc}/h$ run throughout the entire EoR.

On the right panel of Figure \ref{fig:ion_pk}, we plot the power spectrum of the ionization fraction field at $\lambda_{\rm mfp}=1\,{\rm Mpc}/h$ and $\lambda_{\rm mfp}=5\,{\rm Mpc}/h$.
As was expected from the 2D visualizations, the $\lambda_{\rm mfp}=5\,{\rm Mpc}/h$ field has $\sim 3$ times more power on $k<0.5\,{\rm Mpc}^{-1}h$ scales, and half of the power on small scales.
The increase in $\lambda_{\rm mfp}$ induces an almost constant large-scale bias at all ionization levels.
Compared to $M_h$, we see that increasing $\lambda_{\rm mfp}$ has a much stronger effect on the ionization morphology.
Moreover, there is a shift in the peaking scale of $\Delta^2_{\rm ion}$ with $\lambda_{\rm mfp}$.
To correlate the $\lambda_{\rm mfp}$ parameter with the typical sizes of ionized bubbles, we measure the $k_{\rm peak}$ value at which $\Delta^2_{\rm ion}$ peaks, and use $r_{\rm peak}=2 \pi/k_{\rm peak}$ to approximate the characteristic size of ionized bubbles.
The vertical lines in Figure \ref{fig:ion_pk} mark the peaking bubble scales for the ionization fraction $50\%$.
For $\lambda_{\rm mfp}=1\,{\rm Mpc}/h$, the characteristic bubble size is $r_{\rm peak}=8.3\,{\rm Mpc}/h$. 
$\lambda_{\rm mfp}=5\,{\rm Mpc}/h$ corresponds to $r_{\rm peak}=21.6\,{\rm Mpc}/h$.
For more detailed correspondence between $\lambda_{\rm mfp}$ and $r_{\rm peak}$, please refer to the axes of Figure \ref{fig:scale_lambda}.

From the left panel of Figure \ref{fig:ksz_2_reion_params}, we see that when we increase the global mean free path of photons, the peak of the spectra is shifted towards the higher end of $\ell$, corresponding to a larger angular scale subtended by the ionized bubbles. 
At our fiducial $\lambda_{\rm mfp}=3.0\,{\rm Mpc}/h$, the spectra peaks at $\ell=2300$.
Our finding is consistent with previous works \citep[e.g.][]{Gorce2020A&A...640A..90G}, although our $\lambda_{\rm mfp}$ parameter is different from their characterization of the bubble sizes. 
We note that our current model has no spatial and temporal variation of $\lambda_{\rm mfp}$.
In future works the halo mass and redshift dependence of $\lambda_{\rm mfp}$ will
be studied.
The details of where these spectra peak and how they shift with the mean free path will be studied later in Section \ref{sec:scaling}.

\subsection{Scaling of $D^{\rm pkSZ}_{\ell=3000}$ with Reionization Parameters}
\label{sec:scaling}

\begin{figure}[t]
\includegraphics[width=\linewidth]{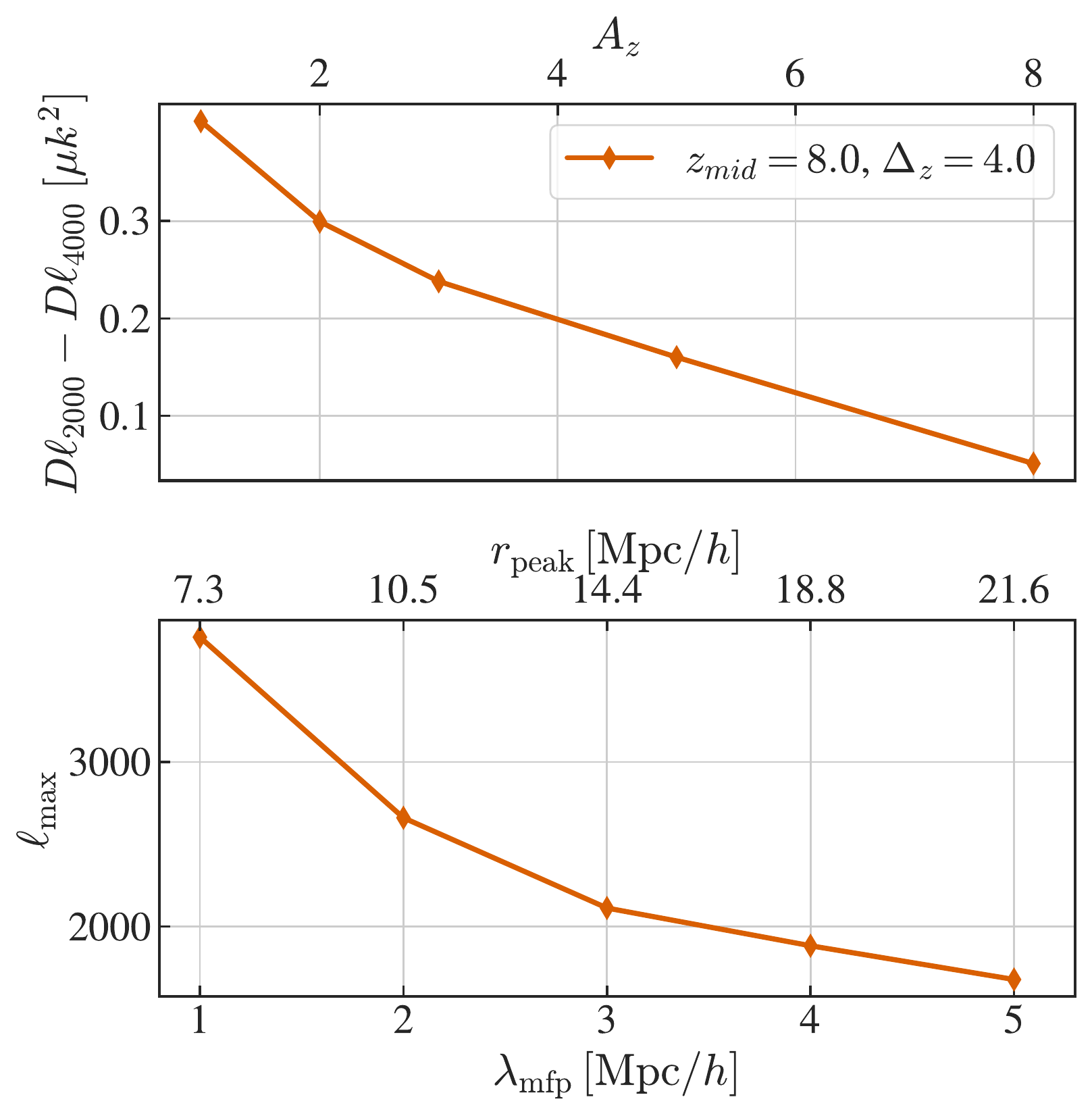} 
\caption{\textbf{\textit{Top:}} Change in the slope of the kSZ power spectrum with the asymmetry of reionization history $A_z$. \textbf{\textit{Bottom:}} Shift in the peak of kSZ spectrum with the mean free path parameter $\lambda_{\rm mfp}$ (\textit{bottom axis}) and the mean bubble sizes (\textit{top axis}).}
\label{fig:scale_lambda}
\end{figure}

\begin{figure*}[t]
\centering
\includegraphics[width=0.87\textwidth]{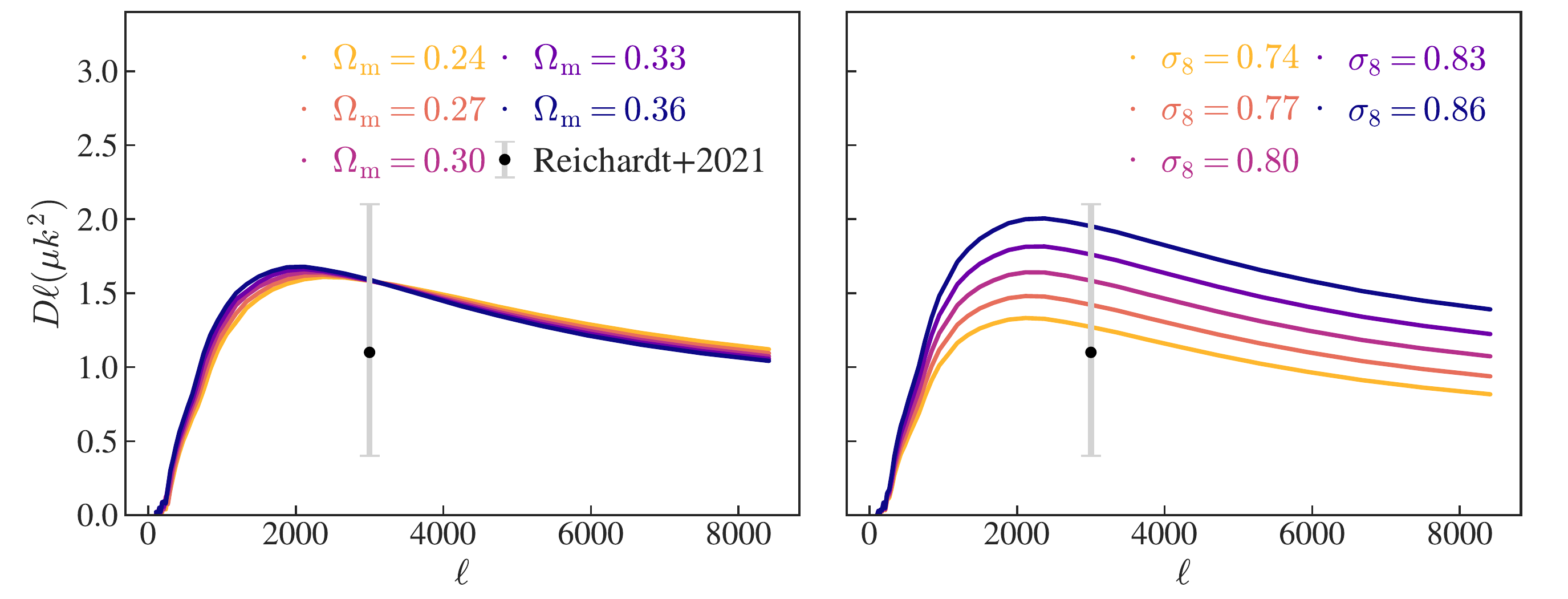}
\caption{Patchy kSZ angular power spectrum with the same reionization history but different cosmological parameters ($\Omega_m$ and $\sigma_8$). While there is complete degeneracy between different $\Omega_m$ values at $\ell=3000$, $\sigma_8$ affects the kSZ amplitude on all scales.} 
\label{fig:cosmo}
\end{figure*}

As was discussed in the previous sections, the amplitude of the kSZ spectrum is most sensitive to $z_{\rm mid}$ and $\Delta_z$, the slope of the spectra is affected by $A_z$, and the peak of the kSZ spectrum is most sensitive to $\lambda_{\text{mfp}}$. 
Now we want to study more quantitatively the dependence of the amplitude and shape of the kSZ spectrum on reionization parameters. 
Note that for the scaling relation study, we always only change one parameter at a time, and keep the other parameters fixed at their fiducial values ($[z_{\rm mid},\Delta_z,A_z,{\rm log}(M_h),\lambda_{\rm mfp}]=[8.0,4.0,3.0,8.0,3.0]$).

In Figure \ref{fig:scale_dz}, we show the amplitude of kSZ at $\ell=3000$ with different $z_{\rm mid}$ and $\Delta_z$ values, respectively. 
From both panels, $D^{\rm pkSZ}_{\ell=3000}$ scales almost linearly with the parameter values. 
For comparison, we also plotted the scaling relation fitted in \citet{Battaglia2013ApJ...776...83B} (Equation (10) in their paper). 
Note that in \citet{Battaglia2013ApJ...776...83B}, the duration is defined to be $\Delta_{z,50}$, and the asymmetry parameter is not measured. 
For comparison with our $\Delta_z$, we assume an asymmetry of $A_z=3$ and use the Weibull function (Equation \ref{eqn:weibull}) to specify the reionization history at $\Delta_z=[2,3,4,5,6]$. Then, for each of these reionization histories, we measure the value of $\Delta_{z,50}$. 
Finally, we input these measured $\Delta_{z,50}$ values into the \citet{Battaglia2013ApJ...776...83B} fits together with a specified $z_{\rm mid}$ value to obtain the green lines.

Comparing with \citet{Battaglia2013ApJ...776...83B}'s power law index of $0.47$ on the duration, our measured $D^{\rm pkSZ}_{\ell=3000}$ has a steeper dependence on $\Delta_z$. 
This is consistent with the findings in \citet{Gorce2020A&A...640A..90G}, although we did not directly show their results as we have different $z_{\rm mid}$ values. 
One possible difference in the scaling is the asymmetrical nature of our reionization histories.
As was shown in Figure \ref{fig:d50_d90}, if the asymmetry of the reionization history is not fixed, the relation between $\Delta_{z,50}$ and $\Delta_{z,90}$ may not be linear. 
Thus \citet{Battaglia2013ApJ...776...83B}'s 0.47 power-law index could result from an increase in asymmetry with the duration under their model.
Another possible explanation is that \citet{Battaglia2013ApJ...776...83B} does not independently control the mean-free path parameter, which degenerates with $\Delta_z$ at $\ell=3000$.
To demonstrate the effect of $\lambda_{\rm mfp}$ on the $\Delta_z$ dependency, on the right panel we show the $D^{\rm pkSZ}_{\ell=3000}-\Delta_z$ relation with $\lambda_{\rm mfp}=1\,{\rm Mpc}/h$ and $\lambda_{\rm mfp}=5\,{\rm Mpc}/h$.
We see that $D^{\rm pkSZ}_{\ell=3000}$ actually peaks around $\lambda_{\rm mfp}=3\,{\rm Mpc}/h$, and the values at $\lambda_{\rm mfp}=1\,{\rm Mpc}/h$ and $\lambda_{\rm mfp}=5\,{\rm Mpc}/h$ are both lower.
In the \citet{Battaglia2013ApJ...776...83B} model, because there is a decrease in $\lambda_{\rm mfp}$ with increased duration \citep[see e.g. Figure 9 of][]{Battaglia2013ApJ...776...81B} since the large-scale bias parameter of the reionization redshift field is fixed \citep[e.g.][]{Trac2021}, the $D^{\rm pkSZ}_{\ell=3000}-\Delta_z$ relation deviates from a linear relationship.

Next, we look at how the asymmetry of reionization and $\lambda_{\text{mfp}}$ affect different aspects of the kSZ spectrum.
In the top panel of Figure \ref{fig:scale_lambda}, we plot the difference between the amplitudes at $\ell=2000$ and $\ell=4000$ as a function of $A_z$.
As we have discussed in Section \ref{sec:asy}, the spectra get flatter as asymmetry rises, and so $D^{\rm pkSZ}_{\ell=2000}-D^{\rm pkSZ}_{\ell=4000}$ falls with larger $A_z$. 
Previously, \citet{Gorce2020A&A...640A..90G} argued that focusing on $D_{\ell=3000}$ is not sufficient to characterize the kSZ signal, especially with various reionization scenarios that lead to asymmetric reionization histories. 
Our result further supports this argument by showing a quantitative scaling between the slope of the kSZ power and the asymmetry of reionization. 

In the bottom panel of Figure \ref{fig:scale_lambda} we plot the location of the kSZ power spectrum peaks as a function of $\lambda_{\text{mfp}}$. 
From the plot we see that $\ell_{\rm max}$ scales as $1/\lambda_{\text{mfp}}$. 
This is expected as $\lambda_{\text{mfp}}$ is correlated with the size of ionized regions, and is in general agreement with Figure 9 in \citet{Gorce2020A&A...640A..90G}. 
To establish a correspondence between our effective mean-free path parameter and the typical ionized bubble sizes at $z_{\rm mid}$, on the top axis we label the peaking scale of the ionizing fraction power spectrum (i.e. the right panel of Figure \ref{fig:ion_pk}).
At our fiducial $\lambda_{\rm mfp}=3\,{\rm Mpc}/h$, the typical bubble size is $14.4\,{\rm Mpc}/h$ (comoving) at $z=8$.

We note that for all the scaling relations shown in this section, we have always fixed all the other reionization parameters and varied one at a time.
This means that all the relations are conditioned, and so one should take caution when using such scaling relations directly to perform parameter constraints.

\subsection{Cosmological Parameters}
\label{sec:cosmo}

\begin{figure}[t]
\centering
\includegraphics[width=0.48\textwidth]{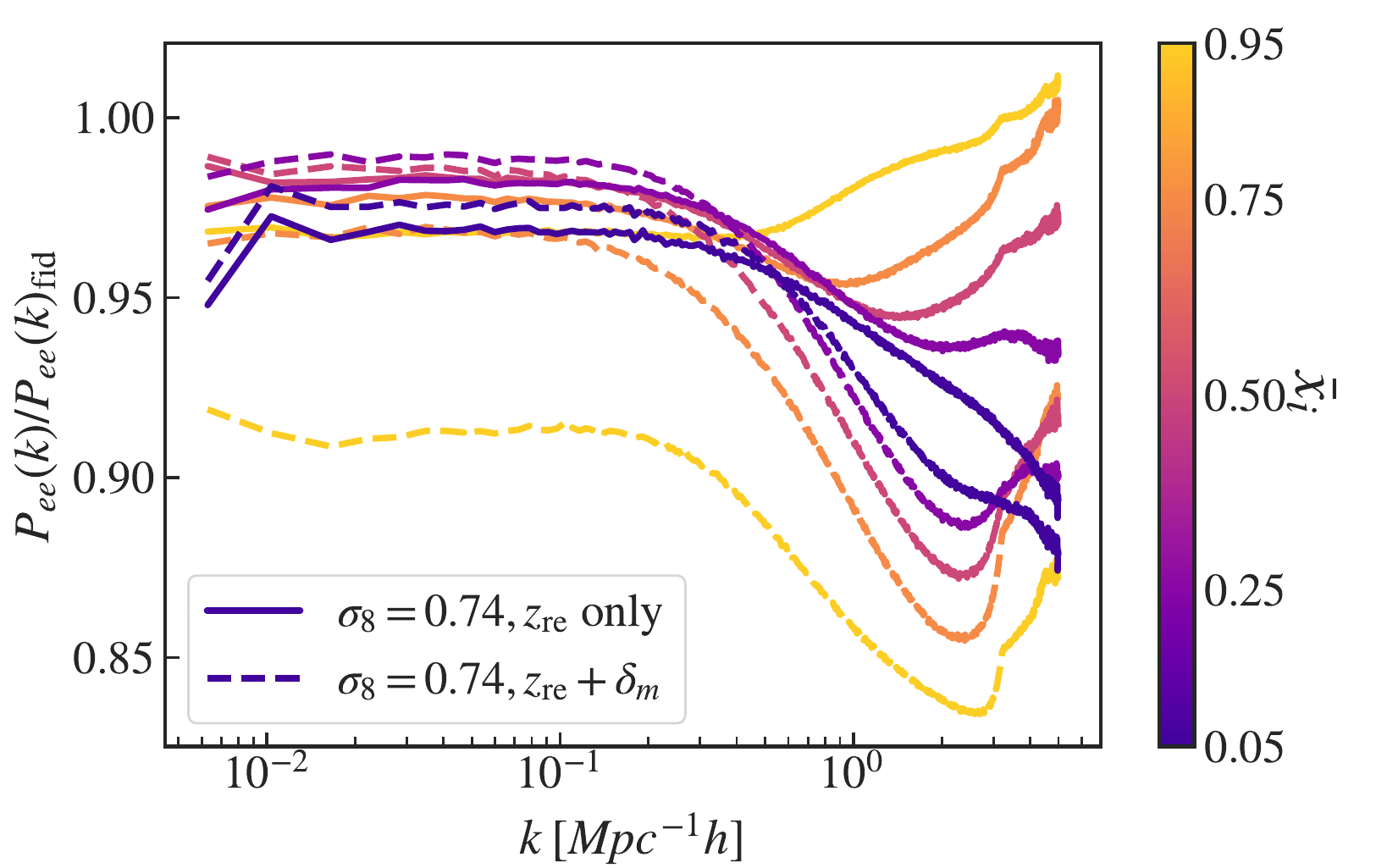}
\caption{The ratio between the electron number density power spectra $P_{ee}(k)$ at $\sigma_8=0.74$ and $\sigma_8=0.8$ (the fiducial value). The the solid curves are generated by changing only the reionization redshift field $z_{\rm re}$ to $\sigma_8=0.74$, while the dashed curves are generated by changing both $z_{\rm re}$ and the matter density field $\delta_m$. The different colors represent the spectra at different ionization levels.} 
\label{fig:ne_s8}
\end{figure}

The process of reionization involves a complicated interplay between cosmology and astrophysical parameters, a large fraction of which remains highly uncertain. 
For instance, a larger $\sigma_8$ could lead to earlier onset of reionization, provided that the nature of the ionizing sources and the photon escape fraction is fixed.
However, there is a lack of comprehensive study on how different cosmologies affect the astrophysics of reionization.
In AMBER, the reionization history and cosmology are modeled independently.
This circumvents the complicated treatment and unknown relation between the two, and allows us to separately analyze the effect of cosmology on the patchy kSZ signal.
In this section, we study the change in the kSZ power spectra when we change two cosmological parameters $\Omega_{\rm m}$ and $\sigma_8$. 

In Figure \ref{fig:cosmo}, we show the change in the patchy kSZ spectra when we vary $\Omega_m$ and $\sigma_8$.
From the left panel, we can see that $D^{\rm pkSZ}_{\ell}$ is only weakly sensitive to the change in $\Omega_m$, especially at $\ell=3000$. 
Larger $\Omega_m$ leads to slightly higher amplitude on $\ell<2000$ scales. 
On the right panel, $\sigma_8$ has more direct effects on the amplitude of the kSZ spectrum: the kSZ power doubles when we change $\sigma_8$ from $0.74$ to $0.86$.
Such effect comes from two different sources. 
Firstly, in linear theory, both the density and velocity fluctuations scale as $\sigma_8^2$. 
Since the kSZ effect measures the momentum fluctuations, we should expect $\sim \sigma_8^4$ contribution from the change in the matter density and velocity fields \citep[also see e.g.][]{Trac2011,Shaw2012}.
Secondly, the change in density contrast and clustering can also influence the reionization redshift field through the radiation intensity. 
Hence, a change in $\sigma_8$ will also affect the morphology of ionized regions at different redshifts.
By comparing with the effects of reionization parameters in Figure \ref{fig:ksz_3_reion_params} and \ref{fig:ksz_2_reion_params}, we see that there are degeneracies between $\sigma_8$, $\Delta_z$, $z_{\rm mid}$ and $\lambda_{\rm mfp}$ in terms of $D^{\rm pkSZ}_{\ell=3000}$.

In order to disentangle the change due to matter density from the change due to the reionization redshift field when varying $\sigma_8$, we show in Figure \ref{fig:ne_s8} the change in $P_{ee}(k,z)$ purely from $z_{\rm re}$, compared with the total change, at different ionization levels.
Here we compare the ratio between $P_{ee}(k,z)$ at $\sigma_8=0.74$ with $P_{ee}(k,z)$ at the fiducial $\sigma_8=0.8$.
For the dashed curve, we simply change the value of $\sigma_8$ in the code, so that both the gas density and the reionization redshift fields are affected.
For the solid curve, we use $\sigma_8=0.74$ to generate the reionization redshift field, while the matter overdensity is kept at $\sigma_8=0.8$.
By comparing the dashed curves at different ionization levels, we see that as the ionization level $\bar{x}_i$ raises, the ratio between the two $P_{ee}$'s drops on all scales. 
Noticeably, only near the end of reionization at $\bar{x}_i=0.95$ does the $P_{\rm ee}$ ratio approach the expected matter power spectrum ratio of $0.85$ on large scales.
At higher ionization levels, the effect of the matter density field is subdominant, especially on large scales, where $>95\%$ of power is retained. 
On small scales ($k>1\,{\rm Mpc}/h$), however, the power drops significantly.
By comparing with the solid curves where only $z_{\rm re}$ varies, we can see that the suppression on small scales still comes from the change in $\delta_m$ instead of $z_{\rm re}$, as the suppression in the solid curves are not as significant.
Therefore, we conclude that changing $\sigma_8$ mostly affects $P_{ee}$ near the end of reionization.
Before the end of reionization, the change in $z_{\rm re}$ has a dominant effect over $\delta_m$ on large scales and only mildly affects $P_{ee}$. On small scales, $P_{ee}$ is mainly affected by the matter density and varies more significantly with $\sigma_8$.

\section{Conclusion}
\label{sec:conclusion}
In this work, we use the new semi-numerical code for reionization AMBER to study the patchy kSZ effect under different reionization scenarios.
We calibrate and test the AMBER predictions against the radiative-transfer RadHydro simulation suite  \citep{Doussot2019ApJ...870...18D}.
We find that at our target resolution of $1\,{\rm Mpc}/h$, AMBER produces electron number density field and kSZ angular power spectra that resemble those from RadHydro at all redshifts.

AMBER explicitly parametrizes the reionization history by the midpoint redshift, duration, and asymmetry parameters. 
By varying the midpoint redshift of reionization, we find that the range $z_{\rm mid}=[6.0,8.9]$ has Thomson optical depth values consistent with the \cite{Planck2018arXiv180706209P} measurements at the $2\sigma$ level. 
We also find that the peaking scale of the kSZ angular power spectrum is not sensitive to the midpoint redshift. 

Then, assuming a value of $z_{\rm mid}=8.0$ consistent with the Planck measurement, and fixing the other parameters at their fiducial values, we find that the amplitude of $D^{\rm pkSZ}_{\ell}$ at $\ell=3000$ scales linearly with the duration of reionization $\Delta_z$.
The resulting $D^{\rm pkSZ}_{\ell=3000}$ values are consistent with the $1\sigma$ measurement from \cite{Reichardt2021ApJ...908..199R} up to $\Delta_z<5.1$ ($\Delta_z$ here encloses redshifts from $5\%$ to $95\%$ reionization). 
This translates to $\Delta_{z,50}<2.0$ under the definition of duration in \cite{Battaglia2013ApJ...776...81B}, assuming a mildly asymmetric reionization history at $A_z=3$. 
Then, allowing for other reionization parameters to vary simultaneously, we find that $\Delta_{z}<12.8$ is the maximum duration consistent with the \cite{Reichardt2021ApJ...908..199R} estimation at the $2\sigma$ level ($\Delta_{z,50}<3.5$ assuming $A_z=8$).
Note that this extreme scenario requires a high asymmetry of the reionization hitory of $A_z>8$, in order for reionization to end before $z=5.5$.
This is in broad agreement with the constraint from \cite{Reichardt2021ApJ...908..199R} of $\Delta_{z,50}<5.4$ ($95\%$ CL) using the \cite{Battaglia2013ApJ...776...83B} model, and the constraint by \cite{Choudhury2021} at $\Delta_{z,50}<2.9$ ($99\%$ CL) using a different semi-numerical model.

Then, by considering reionization histories with different degrees of asymmetry, we find that the kSZ amplitude at $\ell=3000$ is not sensitive to the detailed shape of reionization history beyond redshift and duration.
However, the slope of the kSZ angular power spectrum does depend on the asymmetry.
This is in line with the results shown in \citep[e.g.][]{Park2013ApJ...769...93P,Gorce2020A&A...640A..90G}, and makes constraints on the beginning and end of reionization through patchy kSZ possible if measurements are made at different multipoles.
Nevertheless, we find that constraints on the asymmetry require $\sim 0.1\,\mu k^2$ measurement accuracy of the patchy kSZ power spectrum at various multipoles other than $\ell=3000$.

We also independently control the size of sources through the minimum halo mass ($M_h$), and the relative radiation intensity through the effective mean free path ($\lambda_{\rm mfp}$). 
With a fixed reionization history, the minimum halo mass has little effect on the ionization morphology at a fixed redshift.
Therefore, the amplitude and shape of the kSZ spectrum are only mildly affected by the minimum halo mass $M_h$.
This is in contrast to the results shown in \cite{Paul2021}, and thus a more detailed investigation of which assumptions in our models lead to the differences is needed.
The effective photon mean-free path affects the peaking location of the kSZ power spectrum, and at our fiducial $\lambda_{\rm mfp}=3\,{\rm Mpc}/h$ (fitted to the RadHydro simulations), the spectrum peaks at $\ell \approx 2100$.
Moreover, we explicitly showed that there is a degeneracy between the mean free path $\lambda_{\rm mfp}$ and the duration of reionization in terms of $D^{\rm pkSZ}_{\ell}$ at $\ell=3000$.
A shorter $\lambda_{\rm mfp}$ can lead to a $\sim 10\%$ lower $D^{\rm pkSZ}_{\ell=3000}$ and a flatter slope in the $D^{\rm pkSZ}_{\ell=3000}-\Delta_z$ scaling relation.
This partly explains the steeper power-law scaling relationship we get compared with \cite{Battaglia2013ApJ...776...83B}, as the ionized bubble sizes in their model decrease with a longer duration.

Finally, we study the effect of cosmological parameters $\Omega_m$ and $\sigma_8$ on the patchy kSZ power spectrum under fixed reionization parameters. 
We find that with a fixed reionization history, the kSZ power spectrum does not have noticeable change with $\Omega_m$, especially near $\ell=3000$.
However, $\sigma_8$ affects the overall amplitude of the kSZ power spectrum, which results in a degeneracy between $\sigma_8$ and $\Delta_z$.
On large scales ($k<1\,{\rm Mpc}^{-1}h$), $\sigma_8$ affects the electron number density mainly through the $z_{\rm re}$ field, while on small scales ($k<1\,{\rm Mpc}^{-1}h$), the effect comes from the matter density field.

Even though we have given a rough estimate of the duration of reionization consistent with current observations from ground-based telescopes, such constraints are only a first-order estimation because we only search a 1D parameter space at a time.
To carry out the analysis properly, we need to take into account the correlation between different parameters by marginalizing over other model parameters.
In order to achieve that, we will need a tool to estimate the kSZ spectra faster than what we can achieve with our simulations.

Moreover, in order to separate out the patchy component from the spectrum, one would need a good description of the homogeneous spectrum. Currently, the homogeneous spectrum quoted in \cite{Reichardt2021ApJ...908..199R} comes from \cite{Shaw2012}, but we can use results updated with more recent simulations \citep[e.g.][]{He2021arXiv210704606H} to get a better estimation of how accurately one can recover the patchy signal.

\section*{Acknowledgements}
We thank Marcelo Alvarez and Matthew McQuinn for reading the manuscript and providing important comments and suggestions. H.T.~acknowledges support from the NSF AI Institute:~Planning:~Physics of the Future, NSF PHY2020295.
The simulations were run on the Vera and Bridges-2 clusters at the Pittsburgh Supercomputing Center.

\bibliography{main.bib}{}

\begin{thebibliography}{}
\expandafter\ifx\csname natexlab\endcsname\relax\def\natexlab#1{#1}\fi
\providecommand{\url}[1]{\href{#1}{#1}}
\providecommand{\dodoi}[1]{doi:~\href{http://doi.org/#1}{\nolinkurl{#1}}}
\providecommand{\doeprint}[1]{\href{http://ascl.net/#1}{\nolinkurl{http://ascl.net/#1}}}
\providecommand{\doarXiv}[1]{\href{https://arxiv.org/abs/#1}{\nolinkurl{https://arxiv.org/abs/#1}}}

\bibitem[{{Addison} {et~al.}(2013){Addison}, {Dunkley}, \&
  {Bond}}]{Addison2013}
{Addison}, G.~E., {Dunkley}, J., \& {Bond}, J.~R. 2013, \mnras, 436, 1896,
  \dodoi{10.1093/mnras/stt1703}

\bibitem[{{Alvarez}(2016)}]{Alvarez2016ApJ...824..118A}
{Alvarez}, M.~A. 2016, \apj, 824, 118, \dodoi{10.3847/0004-637X/824/2/118}

\bibitem[{{Alvarez} {et~al.}(2021){Alvarez}, {Ferraro}, {Hill}, {Hlo{\v{z}}ek},
  \& {Ikape}}]{Alvarez2021}
{Alvarez}, M.~A., {Ferraro}, S., {Hill}, J.~C., {Hlo{\v{z}}ek}, R., \& {Ikape},
  M. 2021, \prd, 103, 063518, \dodoi{10.1103/PhysRevD.103.063518}

\bibitem[{{Battaglia} {et~al.}(2013{\natexlab{a}}){Battaglia}, {Natarajan},
  {Trac}, {Cen}, \& {Loeb}}]{Battaglia2013ApJ...776...83B}
{Battaglia}, N., {Natarajan}, A., {Trac}, H., {Cen}, R., \& {Loeb}, A.
  2013{\natexlab{a}}, \apj, 776, 83, \dodoi{10.1088/0004-637X/776/2/83}

\bibitem[{{Battaglia} {et~al.}(2013{\natexlab{b}}){Battaglia}, {Trac}, {Cen},
  \& {Loeb}}]{Battaglia2013ApJ...776...81B}
{Battaglia}, N., {Trac}, H., {Cen}, R., \& {Loeb}, A. 2013{\natexlab{b}}, \apj,
  776, 81, \dodoi{10.1088/0004-637X/776/2/81}

\bibitem[{{Becker} {et~al.}(2015){Becker}, {Bolton}, {Madau}, {Pettini},
  {Ryan-Weber}, \& {Venemans}}]{Becker2015MNRAS.447.3402B}
{Becker}, G.~D., {Bolton}, J.~S., {Madau}, P., {et~al.} 2015, \mnras, 447,
  3402, \dodoi{10.1093/mnras/stu2646}

\bibitem[{{Bond} {et~al.}(1991){Bond}, {Cole}, {Efstathiou}, \&
  {Kaiser}}]{Bond1991ApJ...379..440B}
{Bond}, J.~R., {Cole}, S., {Efstathiou}, G., \& {Kaiser}, N. 1991, \apj, 379,
  440, \dodoi{10.1086/170520}

\bibitem[{{Bouwens} {et~al.}(2015){Bouwens}, {Illingworth}, {Oesch}, {Trenti},
  {Labb{\'e}}, {Bradley}, {Carollo}, {van Dokkum}, {Gonzalez}, {Holwerda},
  {Franx}, {Spitler}, {Smit}, \& {Magee}}]{Bouwens2015ApJ...803...34B}
{Bouwens}, R.~J., {Illingworth}, G.~D., {Oesch}, P.~A., {et~al.} 2015, \apj,
  803, 34, \dodoi{10.1088/0004-637X/803/1/34}

\bibitem[{{Carlstrom} {et~al.}(2002){Carlstrom}, {Holder}, \&
  {Reese}}]{Carlstrom2002}
{Carlstrom}, J.~E., {Holder}, G.~P., \& {Reese}, E.~D. 2002, \araa, 40, 643,
  \dodoi{10.1146/annurev.astro.40.060401.093803}

\bibitem[{{Chen} {et~al.}(2020){Chen}, {Doussot}, {Trac}, \&
  {Cen}}]{Chen2020ApJ...905..132C}
{Chen}, N., {Doussot}, A., {Trac}, H., \& {Cen}, R. 2020, \apj, 905, 132,
  \dodoi{10.3847/1538-4357/abc890}

\bibitem[{{Choudhury} {et~al.}(2021){Choudhury}, {Mukherjee}, \&
  {Paul}}]{Choudhury2021}
{Choudhury}, T.~R., {Mukherjee}, S., \& {Paul}, S. 2021, \mnras, 501, L7,
  \dodoi{10.1093/mnrasl/slaa185}

\bibitem[{{de Belsunce} {et~al.}(2021){de Belsunce}, {Gratton}, {Coulton}, \&
  {Efstathiou}}]{deBelsunce2021MNRAS.507.1072D}
{de Belsunce}, R., {Gratton}, S., {Coulton}, W., \& {Efstathiou}, G. 2021,
  \mnras, 507, 1072, \dodoi{10.1093/mnras/stab2215}

\bibitem[{{DeBoer} {et~al.}(2017){DeBoer}, {Parsons}, {Aguirre}, {Alexander},
  {Ali}, {Beardsley}, {Bernardi}, {Bowman}, {Bradley}, {Carilli}, {Cheng}, {de
  Lera Acedo}, {Dillon}, {Ewall-Wice}, {Fadana}, {Fagnoni}, {Fritz},
  {Furlanetto}, {Glendenning}, {Greig}, {Grobbelaar}, {Hazelton}, {Hewitt},
  {Hickish}, {Jacobs}, {Julius}, {Kariseb}, {Kohn}, {Lekalake}, {Liu}, {Loots},
  {MacMahon}, {Malan}, {Malgas}, {Maree}, {Martinot}, {Mathison}, {Matsetela},
  {Mesinger}, {Morales}, {Neben}, {Patra}, {Pieterse}, {Pober}, {Razavi-Ghods},
  {Ringuette}, {Robnett}, {Rosie}, {Sell}, {Smith}, {Syce}, {Tegmark},
  {Thyagarajan}, {Williams}, \& {Zheng}}]{HERA2017}
{DeBoer}, D.~R., {Parsons}, A.~R., {Aguirre}, J.~E., {et~al.} 2017, \pasp, 129,
  045001, \dodoi{10.1088/1538-3873/129/974/045001}

\bibitem[{{Doussot} {et~al.}(2019){Doussot}, {Trac}, \&
  {Cen}}]{Doussot2019ApJ...870...18D}
{Doussot}, A., {Trac}, H., \& {Cen}, R. 2019, \apj, 870, 18,
  \dodoi{10.3847/1538-4357/aaef75}

\bibitem[{{Dunkley} {et~al.}(2013){Dunkley}, {Calabrese}, {Sievers}, {Addison},
  {Battaglia}, {Battistelli}, {Bond}, {Das}, {Devlin}, {D{\"u}nner}, {Fowler},
  {Gralla}, {Hajian}, {Halpern}, {Hasselfield}, {Hincks}, {Hlozek}, {Hughes},
  {Irwin}, {Kosowsky}, {Louis}, {Marriage}, {Marsden}, {Menanteau}, {Moodley},
  {Niemack}, {Nolta}, {Page}, {Partridge}, {Sehgal}, {Spergel}, {Staggs},
  {Switzer}, {Trac}, \& {Wollack}}]{Dunkley2013JCAP...07..025D}
{Dunkley}, J., {Calabrese}, E., {Sievers}, J., {et~al.} 2013, \jcap, 2013, 025,
  \dodoi{10.1088/1475-7516/2013/07/025}

\bibitem[{{Ferraro} \& {Smith}(2018)}]{Ferraro2018}
{Ferraro}, S., \& {Smith}, K.~M. 2018, \prd, 98, 123519,
  \dodoi{10.1103/PhysRevD.98.123519}

\bibitem[{{Finkelstein} {et~al.}(2015){Finkelstein}, {Ryan}, {Papovich},
  {Dickinson}, {Song}, {Somerville}, {Ferguson}, {Salmon}, {Giavalisco},
  {Koekemoer}, {Ashby}, {Behroozi}, {Castellano}, {Dunlop}, {Faber}, {Fazio},
  {Fontana}, {Grogin}, {Hathi}, {Jaacks}, {Kocevski}, {Livermore}, {McLure},
  {Merlin}, {Mobasher}, {Newman}, {Rafelski}, {Tilvi}, \&
  {Willner}}]{Finkelstein2015ApJ...810...71F}
{Finkelstein}, S.~L., {Ryan}, Jr., R.~E., {Papovich}, C., {et~al.} 2015, \apj,
  810, 71, \dodoi{10.1088/0004-637X/810/1/71}

\bibitem[{{Furlanetto} {et~al.}(2004){Furlanetto}, {Zaldarriaga}, \&
  {Hernquist}}]{Furlanetto2004ApJ...613....1F}
{Furlanetto}, S.~R., {Zaldarriaga}, M., \& {Hernquist}, L. 2004, \apj, 613, 1,
  \dodoi{10.1086/423025}

\bibitem[{{George} {et~al.}(2015){George}, {Reichardt}, {Aird}, {Benson},
  {Bleem}, {Carlstrom}, {Chang}, {Cho}, {Crawford}, {Crites}, {de Haan},
  {Dobbs}, {Dudley}, {Halverson}, {Harrington}, {Holder}, {Holzapfel}, {Hou},
  {Hrubes}, {Keisler}, {Knox}, {Lee}, {Leitch}, {Lueker}, {Luong-Van},
  {McMahon}, {Mehl}, {Meyer}, {Millea}, {Mocanu}, {Mohr}, {Montroy}, {Padin},
  {Plagge}, {Pryke}, {Ruhl}, {Schaffer}, {Shaw}, {Shirokoff}, {Spieler},
  {Staniszewski}, {Stark}, {Story}, {van Engelen}, {Vanderlinde}, {Vieira},
  {Williamson}, \& {Zahn}}]{George2015ApJ...799..177G}
{George}, E.~M., {Reichardt}, C.~L., {Aird}, K.~A., {et~al.} 2015, \apj, 799,
  177, \dodoi{10.1088/0004-637X/799/2/177}

\bibitem[{{Glazer} {et~al.}(2018){Glazer}, {Rau}, \&
  {Trac}}]{Glazer2018RNAAS...2c.135G}
{Glazer}, D., {Rau}, M.~M., \& {Trac}, H. 2018, Research Notes of the American
  Astronomical Society, 2, 135, \dodoi{10.3847/2515-5172/aad68a}

\bibitem[{{Gorce} {et~al.}(2020){Gorce}, {Ili{\'c}}, {Douspis}, {Aubert}, \&
  {Langer}}]{Gorce2020A&A...640A..90G}
{Gorce}, A., {Ili{\'c}}, S., {Douspis}, M., {Aubert}, D., \& {Langer}, M. 2020,
  \aap, 640, A90, \dodoi{10.1051/0004-6361/202038170}

\bibitem[{{G{\'o}rski} {et~al.}(2005){G{\'o}rski}, {Hivon}, {Banday},
  {Wandelt}, {Hansen}, {Reinecke}, \& {Bartelmann}}]{Gorski2005}
{G{\'o}rski}, K.~M., {Hivon}, E., {Banday}, A.~J., {et~al.} 2005, \apj, 622,
  759, \dodoi{10.1086/427976}

\bibitem[{{He} {et~al.}(2021){He}, {Trac}, \& {Gnedin}}]{He2021arXiv210704606H}
{He}, Y., {Trac}, H., \& {Gnedin}, N.~Y. 2021, arXiv e-prints,
  arXiv:2107.04606.
\newblock \doarXiv{2107.04606}

\bibitem[{{Iliev} {et~al.}(2007){Iliev}, {Pen}, {Bond}, {Mellema}, \&
  {Shapiro}}]{Iliev2007ApJ...660..933I}
{Iliev}, I.~T., {Pen}, U.-L., {Bond}, J.~R., {Mellema}, G., \& {Shapiro}, P.~R.
  2007, \apj, 660, 933, \dodoi{10.1086/513687}

\bibitem[{{Keating} {et~al.}(2019){Keating}, {Weinberger}, {Kulkarni},
  {Haehnelt}, {Chardin}, \& {Aubert}}]{Keating2019arXiv190512640K}
{Keating}, L.~C., {Weinberger}, L.~H., {Kulkarni}, G., {et~al.} 2019, arXiv
  e-prints, arXiv:1905.12640.
\newblock \doarXiv{1905.12640}

\bibitem[{{Koopmans} {et~al.}(2015){Koopmans}, {Pritchard}, {Mellema},
  {Aguirre}, {Ahn}, {Barkana}, {van Bemmel}, {Bernardi}, {Bonaldi}, {Briggs},
  {de Bruyn}, {Chang}, {Chapman}, {Chen}, {Ciardi}, {Dayal}, {Ferrara},
  {Fialkov}, {Fiore}, {Ichiki}, {Illiev}, {Inoue}, {Jelic}, {Jones}, {Lazio},
  {Maio}, {Majumdar}, {Mack}, {Mesinger}, {Morales}, {Parsons}, {Pen},
  {Santos}, {Schneider}, {Semelin}, {de Souza}, {Subrahmanyan}, {Takeuchi},
  {Vedantham}, {Wagg}, {Webster}, {Wyithe}, {Datta}, \& {Trott}}]{SKA2015}
{Koopmans}, L., {Pritchard}, J., {Mellema}, G., {et~al.} 2015, in Advancing
  Astrophysics with the Square Kilometre Array (AASKA14), 1

\bibitem[{{McQuinn} {et~al.}(2005){McQuinn}, {Furlanetto}, {Hernquist}, {Zahn},
  \& {Zaldarriaga}}]{McQuinn2005}
{McQuinn}, M., {Furlanetto}, S.~R., {Hernquist}, L., {Zahn}, O., \&
  {Zaldarriaga}, M. 2005, \apj, 630, 643, \dodoi{10.1086/432049}

\bibitem[{{Mesinger} {et~al.}(2011){Mesinger}, {Furlanetto}, \&
  {Cen}}]{Mesinger2011MNRAS.411..955M}
{Mesinger}, A., {Furlanetto}, S., \& {Cen}, R. 2011, \mnras, 411, 955,
  \dodoi{10.1111/j.1365-2966.2010.17731.x}

\bibitem[{{Mesinger} {et~al.}(2012){Mesinger}, {McQuinn}, \&
  {Spergel}}]{Mesinger2012MNRAS.422.1403M}
{Mesinger}, A., {McQuinn}, M., \& {Spergel}, D.~N. 2012, \mnras, 422, 1403,
  \dodoi{10.1111/j.1365-2966.2012.20713.x}

\bibitem[{{Ostriker} \& {Vishniac}(1986)}]{Ostriker1986ApJ...306L..51O}
{Ostriker}, J.~P., \& {Vishniac}, E.~T. 1986, \apjl, 306, L51,
  \dodoi{10.1086/184704}

\bibitem[{{Pagano} {et~al.}(2020){Pagano}, {Delouis}, {Mottet}, {Puget}, \&
  {Vibert}}]{Pagano2020A&A...635A..99P}
{Pagano}, L., {Delouis}, J.~M., {Mottet}, S., {Puget}, J.~L., \& {Vibert}, L.
  2020, \aap, 635, A99, \dodoi{10.1051/0004-6361/201936630}

\bibitem[{{Park} {et~al.}(2013){Park}, {Shapiro}, {Komatsu}, {Iliev}, {Ahn}, \&
  {Mellema}}]{Park2013ApJ...769...93P}
{Park}, H., {Shapiro}, P.~R., {Komatsu}, E., {et~al.} 2013, \apj, 769, 93,
  \dodoi{10.1088/0004-637X/769/2/93}

\bibitem[{{Paul} {et~al.}(2021){Paul}, {Mukherjee}, \& {Choudhury}}]{Paul2021}
{Paul}, S., {Mukherjee}, S., \& {Choudhury}, T.~R. 2021, \mnras, 500, 232,
  \dodoi{10.1093/mnras/staa3221}

\bibitem[{{Planck Collaboration} {et~al.}(2018){Planck Collaboration},
  {Aghanim}, {Akrami}, {Ashdown}, {Aumont}, {Baccigalupi}, {Ballardini},
  {Banday}, {Barreiro}, \& {Bartolo}}]{Planck2018arXiv180706209P}
{Planck Collaboration}, {Aghanim}, N., {Akrami}, Y., {et~al.} 2018, arXiv
  e-prints, arXiv:1807.06209.
\newblock \doarXiv{1807.06209}

\bibitem[{{Reichardt} {et~al.}(2021){Reichardt}, {Patil}, {Ade}, {Anderson},
  {Austermann}, {Avva}, {Baxter}, {Beall}, {Bender}, {Benson}, {Bianchini},
  {Bleem}, {Carlstrom}, {Chang}, {Chaubal}, {Chiang}, {Chou}, {Citron},
  {Moran}, {Crawford}, {Crites}, {de Haan}, {Dobbs}, {Everett}, {Gallicchio},
  {George}, {Gilbert}, {Gupta}, {Halverson}, {Harrington}, {Henning}, {Hilton},
  {Holder}, {Holzapfel}, {Hrubes}, {Huang}, {Hubmayr}, {Irwin}, {Knox}, {Lee},
  {Li}, {Lowitz}, {Luong-Van}, {McMahon}, {Mehl}, {Meyer}, {Millea}, {Mocanu},
  {Mohr}, {Montgomery}, {Nadolski}, {Natoli}, {Nibarger}, {Noble}, {Novosad},
  {Omori}, {Padin}, {Pryke}, {Ruhl}, {Saliwanchik}, {Sayre}, {Schaffer},
  {Shirokoff}, {Sievers}, {Smecher}, {Spieler}, {Staniszewski}, {Stark},
  {Tucker}, {Vanderlinde}, {Veach}, {Vieira}, {Wang}, {Whitehorn},
  {Williamson}, {Wu}, \& {Yefremenko}}]{Reichardt2021ApJ...908..199R}
{Reichardt}, C.~L., {Patil}, S., {Ade}, P.~A.~R., {et~al.} 2021, \apj, 908,
  199, \dodoi{10.3847/1538-4357/abd407}

\bibitem[{{Shaw} {et~al.}(2012){Shaw}, {Rudd}, \& {Nagai}}]{Shaw2012}
{Shaw}, L.~D., {Rudd}, D.~H., \& {Nagai}, D. 2012, \apj, 756, 15,
  \dodoi{10.1088/0004-637X/756/1/15}

\bibitem[{Smith \& Ferraro(2017)}]{Smith}
Smith, K.~M., \& Ferraro, S. 2017, Phys. Rev. Lett., 119, 021301,
  \dodoi{10.1103/PhysRevLett.119.021301}

\bibitem[{{Spergel} {et~al.}(2015){Spergel}, {Gehrels}, {Baltay}, {Bennett},
  {Breckinridge}, {Donahue}, {Dressler}, {Gaudi}, {Greene}, {Guyon}, {Hirata},
  {Kalirai}, {Kasdin}, {Macintosh}, {Moos}, {Perlmutter}, {Postman},
  {Rauscher}, {Rhodes}, {Wang}, {Weinberg}, {Benford}, {Hudson}, {Jeong},
  {Mellier}, {Traub}, {Yamada}, {Capak}, {Colbert}, {Masters}, {Penny},
  {Savransky}, {Stern}, {Zimmerman}, {Barry}, {Bartusek}, {Carpenter}, {Cheng},
  {Content}, {Dekens}, {Demers}, {Grady}, {Jackson}, {Kuan}, {Kruk}, {Melton},
  {Nemati}, {Parvin}, {Poberezhskiy}, {Peddie}, {Ruffa}, {Wallace}, {Whipple},
  {Wollack}, \& {Zhao}}]{WFIRST2015}
{Spergel}, D., {Gehrels}, N., {Baltay}, C., {et~al.} 2015, arXiv e-prints,
  arXiv:1503.03757.
\newblock \doarXiv{1503.03757}

\bibitem[{{Sunyaev} \& {Zeldovich}(1980)}]{Sunyaev1980ARA&A..18..537S}
{Sunyaev}, R.~A., \& {Zeldovich}, I.~B. 1980, \araa, 18, 537,
  \dodoi{10.1146/annurev.aa.18.090180.002541}

\bibitem[{{Tashiro} {et~al.}(2011){Tashiro}, {Aghanim}, {Langer}, {Douspis},
  {Zaroubi}, \& {Jeli{\'c}}}]{Tashiro2011}
{Tashiro}, H., {Aghanim}, N., {Langer}, M., {et~al.} 2011, \mnras, 414, 3424,
  \dodoi{10.1111/j.1365-2966.2011.18644.x}

\bibitem[{{Trac} {et~al.}(2011){Trac}, {Bode}, \& {Ostriker}}]{Trac2011}
{Trac}, H., {Bode}, P., \& {Ostriker}, J.~P. 2011, \apj, 727, 94,
  \dodoi{10.1088/0004-637X/727/2/94}

\bibitem[{{Trac} {et~al.}(2008){Trac}, {Cen}, \&
  {Loeb}}]{Trac2008ApJ...689L..81T}
{Trac}, H., {Cen}, R., \& {Loeb}, A. 2008, \apjl, 689, L81,
  \dodoi{10.1086/595678}

\bibitem[{{Trac} {et~al.}(2015){Trac}, {Cen}, \&
  {Mansfield}}]{Trac2015ApJ...813...54T}
{Trac}, H., {Cen}, R., \& {Mansfield}, P. 2015, \apj, 813, 54,
  \dodoi{10.1088/0004-637X/813/1/54}

\bibitem[{{Trac} {et~al.}(2021){Trac}, {Chen}, {Holst}, {Alvarez}, \&
  {Cen}}]{Trac2021}
{Trac}, H., {Chen}, N., {Holst}, I., {Alvarez}, M.~A., \& {Cen}, R. 2021, arXiv
  e-prints, arXiv:2109.10375.
\newblock \doarXiv{2109.10375}

\bibitem[{{Weibull}(1951)}]{Weibull1951JAM....18..293W}
{Weibull}, W. 1951, Journal of Applied Mechanics, 18, 293

\bibitem[{{Windhorst} {et~al.}(2006){Windhorst}, {Cohen}, {Jansen},
  {Conselice}, \& {Yan}}]{JWST2006}
{Windhorst}, R.~A., {Cohen}, S.~H., {Jansen}, R.~A., {Conselice}, C., \& {Yan},
  H. 2006, \nar, 50, 113, \dodoi{10.1016/j.newar.2005.11.018}

\bibitem[{{Zahn} {et~al.}(2007){Zahn}, {Lidz}, {McQuinn}, {Dutta}, {Hernquist},
  {Zaldarriaga}, \& {Furlanetto}}]{Zahn2007ApJ...654...12Z}
{Zahn}, O., {Lidz}, A., {McQuinn}, M., {et~al.} 2007, \apj, 654, 12,
  \dodoi{10.1086/509597}

\bibitem[{{Zahn} {et~al.}(2011){Zahn}, {Mesinger}, {McQuinn}, {Trac}, {Cen}, \&
  {Hernquist}}]{Zahn2011MNRAS.414..727Z}
{Zahn}, O., {Mesinger}, A., {McQuinn}, M., {et~al.} 2011, \mnras, 414, 727,
  \dodoi{10.1111/j.1365-2966.2011.18439.x}

\bibitem[{{Zahn} {et~al.}(2012){Zahn}, {Reichardt}, {Shaw}, {Lidz}, {Aird},
  {Benson}, {Bleem}, {Carlstrom}, {Chang}, {Cho}, {Crawford}, {Crites}, {de
  Haan}, {Dobbs}, {Dor{\'e}}, {Dudley}, {George}, {Halverson}, {Holder},
  {Holzapfel}, {Hoover}, {Hou}, {Hrubes}, {Joy}, {Keisler}, {Knox}, {Lee},
  {Leitch}, {Lueker}, {Luong-Van}, {McMahon}, {Mehl}, {Meyer}, {Millea},
  {Mohr}, {Montroy}, {Natoli}, {Padin}, {Plagge}, {Pryke}, {Ruhl}, {Schaffer},
  {Shirokoff}, {Spieler}, {Staniszewski}, {Stark}, {Story}, {van Engelen},
  {Vanderlinde}, {Vieira}, \& {Williamson}}]{Zahn2012ApJ...756...65Z}
{Zahn}, O., {Reichardt}, C.~L., {Shaw}, L., {et~al.} 2012, \apj, 756, 65,
  \dodoi{10.1088/0004-637X/756/1/65}

\bibitem[{{Zeldovich} \& {Sunyaev}(1969)}]{Zeldovich1969}
{Zeldovich}, Y.~B., \& {Sunyaev}, R.~A. 1969, \apss, 4, 301,
  \dodoi{10.1007/BF00661821}

\bibitem[{{Zhang} {et~al.}(2004){Zhang}, {Pen}, \& {Trac}}]{Zhang2004}
{Zhang}, P., {Pen}, U.-L., \& {Trac}, H. 2004, \mnras, 347, 1224,
  \dodoi{10.1111/j.1365-2966.2004.07298.x}

\end{thebibliography}
\bibliographystyle{aasjournal}

\end{document}